\documentclass[traditabstract]{aa} 
\usepackage{graphicx}
\usepackage{txfonts}

\def\la{\raise.5ex\hbox{$<$}\kern-.8em\lower 1mm\hbox{$\sim$}}
\def\ma{\raise.5ex\hbox{$>$}\kern-.8em\lower 1mm\hbox{$\sim$}}

\def\msol{M$_{\odot}$ }

\def\kms{$\rm km\, s^{-1}$}
\def\cm3{$\rm cm^{-3}$}
\def\Ts{$\rm T_{*}$~}
\def\Vs{$\rm V_{s}$~}
\def\n0{$\rm n_{0}$}
\def\B0{$\rm B_{0}$}
\def\ne{$\rm n_{e}$~}

\def\Te{$\rm T_{e}$}

\def\erg{$\rm erg\, cm^{-2}\, s^{-1}$}
\def\mum{$\mu$m~}

\def\L12{L$_{12\mu m}$~}
\def\F12{F$_{12\mu m}$~}
\def\agr{a$_{gr}$}
\def\Hb{H${\beta}$~}
\def\Ha{H${\alpha}$~}

\def\Ly{Ly$\alpha$~}

\def\Moy{M$_{\odot}$ yr$^{-1}$}

\def\RO3{R$_{[OIII]}$}

\begin{document}

   \title{Observed and predicted \Ly and   UV lines\\ for a sample of galaxies at redshifts z$<$3.7
}


   \author{M. Contini \inst{1,2}
}

   \institute{Dipartimento di Fisica e Astronomia, University of Padova, Vicolo dell'Osservatorio 2. I-35133 Padova, Italy
         \and
             School of Physics and Astronomy, Tel Aviv University, Tel Aviv 69978, Israel\\
}

   \date{Received }


  \abstract{
We  explore the origin of  the observed \Ly  and other  UV lines  from  galaxies  at z $<$ 3.7
by detailed modelling of the spectra.
The  objects  are chosen among those  showing    a) UV - optical - near-IR lines, b)
 only  UV lines
 and c) those  showing    \Ly  in the UV and  a few  optical lines.
We  also present  UV line  predictions for   a sample of galaxies   in  the
  0.0686$<$z$<$0.8829 range. 
The sample of galaxies  including  \Ly observations in their spectra does not  show
peculiar physical conditions of the emitting gas, nor abnormal element abundances.
However, the high velocity (\Vs$\geq$ 1000 \kms) component of the emitting gas is accompanied by relatively low
preshock densities (\n0$\sim$ 100-400 \cm3) leading in some cases to broad forbidden lines.
Some spectra are best reproduced by shock dominated models in which the photoionizing source is hidden
or absent.
Within more than 50 galaxies modelled in this work, only  a few  spectra from galaxies at z$\geq$2.5
 correspond to  a starburst temperature \Ts$>$  10$^5$ K,
similar to that  found in galaxies showing some  activity.  

}


\titlerunning{\Ly and UV line models for galaxies at z$<$3.7}

   \maketitle

\section{Introduction}

 \Ly is one of the strongest lines in the UV because H is the most abundant element 
and  because corresponding  to the 1S-2P transition.   
  \Ly $\lambda$1215  line  in  the far-UV is now   reported by observations  of galaxies
at high redshifts 
 (e.g. Humphrey et al 2008,  Dawson et al 2003,  Stern et al 2002,  Erb et al 2010,  Fosbury et al 2003,
 Vernet et al 2000, Dey et al 2005,  Norman et al 2002, 
 Finkelstein et al 2011 etc).
The   line flux is absorbed by dust, depending on   the
environments of the \Ly emitting nebula.
The grains embedded
into the gaseous clouds or distributed throughout the  photon's  way to Earth can lead to a significant
drop of the flux (Finkelstein et al 2011). The escape probability was treated by Atek et al (2014)
who  confirmed that the \Ly escape fraction depends on the dust extinction,
but the correlation does not follow the expected curve for a simple dust attenuation. They  claim
that a higher attenuation can be attributed also to a scattering process
and  that the strength of \Ly and the escape fraction appear unrelated to the galaxy metallicity.

Supernova (SN) events throughout  galaxies
lead to specific shocks  in the regions surrounding the SN explosion (Heng \& Sunyaev 2008) and
close to outbursts  in novae and symbiotic stars (Contini et al 2009).
Heng \& Sunyaev (2008) 
 claim that charge transfer  reactions between hydrogen atoms and protons in
collisionless shocks of
SN  remnants  produce broad Balmer, Lyman, and other hydrogen lines with increased \Ly/\Hb.

Generally, photoionization by active galactic nuclei (AGN) or by  starbursts  (SB) is the main 
 heating and ionizing   mechanism of  gas  throughout the galaxy. 
When  no photoionization source is seen, and much of the energy is 
provided by the conversion of kinetic energy  of motion into heat, e.g.
when  fast moving matter collides with ambient ISM gas,  
ionization and thermal energy are released. They will be  partly
radiated as recombination and collisional line  emission (Osterbrock 1989).

It was suggested that high redshift  galaxies originate from merging (e.g. Ryan et al 2008). Therefore strong shocks
could appear in some regions of the product galaxy, as  e.g. in the local galaxy NGC 3393
(Contini 2012).
Yajima et al (2012) proposed a formation model from major mergers of gas rich
galaxies emitting the \Ly line. They claim that  at z$\geq$ 2 the  merger rate is higher
in dense region and the progenitors are more gas rich.
Lyman break technique could  be employed to explore galaxies at
0.8 $\leq$ z $\leq$ 2.5 (Oteo et al. 2002).
Merging galaxies   emit  \Ly as well as many other lines.
Observations of  UV lines  are now available from  galaxies  at high z, in particular 
 NV 1240, CIV 1550, HeII 1640, etc. and eventually OVI 1034.
They are emitted by  high ionization-level ions  that are  easily  produced in 
shock heated gas corresponding to  suitable  velocities.
High ionization-level lines are  strong from  C, N, O, Ne, etc.
in  the IR, optical, and UV frequency ranges. Most of the UV lines (e.g. OIV, OV, NV)  are permitted,
while  e.g. [NV], [OV], etc  in the optical and IR are forbidden  and thus  collisionally deexcited at high densities.. 
Permitted lines  originate from recombination  and   depend on  radiative processes.
Both radiation and  collisional processes should be considered in the calculation of  line
   spectra rich enough to constrain the models.
Composite models (shock +photoionization) were
 adopted to explain the spectra from  local merging galaxies (e.g. NGC 7212, NGC 3393, NGC 6240, Arp 220,
Contini 2013a and references therein) and  were used  in the  spectral line analysis of galaxies at a  relatively high z
 (Contini 2014a and references therein.)

In this paper we  explore the origin of  the observed \Ly  and other lines  in the UV from  galaxies  at z $<$ 3.7
by detailed modelling of the spectra. 
 Our aim is to find out not only the gas physical conditions and element abundances throughout the
emitting clouds,  but to determine the nature of the radiation source, such as a AGN, a SB
or only shocks.
The  objects  were chosen among those  showing   a)  UV - optical - near-IR lines 
 (Humphrey et al,  Dawson et al,  Stern et al,  Erb et al   and 
 Fosbury et al),
b) only  UV lines (Vernet et al, Dey et al  and  Norman et al) 
 and c) those  showing    \Ly  in the UV and  a few  optical lines (Finkelstein et al).
We  also present  UV line  predictions for   a sample of galaxies  (Ly et al 2014) in  the
 0.0686$\leq$z$\leq$0.8829 range, where the physical conditions of the emitting gas  are   
constrained by the observed optical lines, in particular  [OIII]4363. 
 The total sample  which includes galaxies in
an all-sky radio survey, 
 hard X-ray emitting sources, a type 2 QSR identified as a X-ray source, 
a type 2 AGN at high z (3.7),  a very large spacially extended \Ly emitting nebula,
objects in  a multi-band
imaging survey for Lyman break galaxies,  in a slitless spectroscopic survey for
LAEs, etc. is rather heterogeneous. 
However, the observed spectra from high z  objects accounting for the  \Ly  line   and  for enough
lines  constraining  the models,
are few, so  we gathered them  from different  classes of objects.

\begin{figure*}
\includegraphics[width=6.0cm]{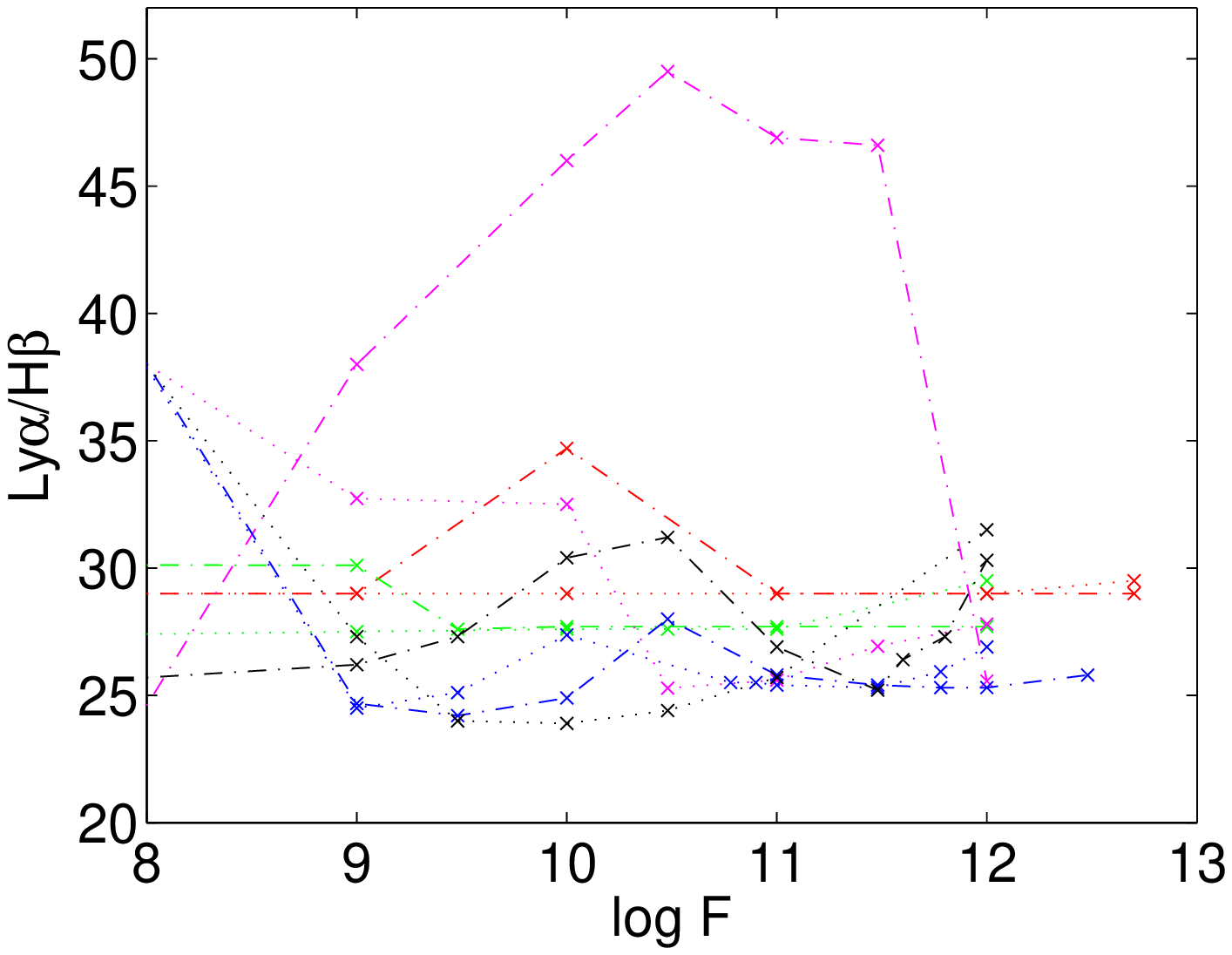}
\includegraphics[width=6.0cm]{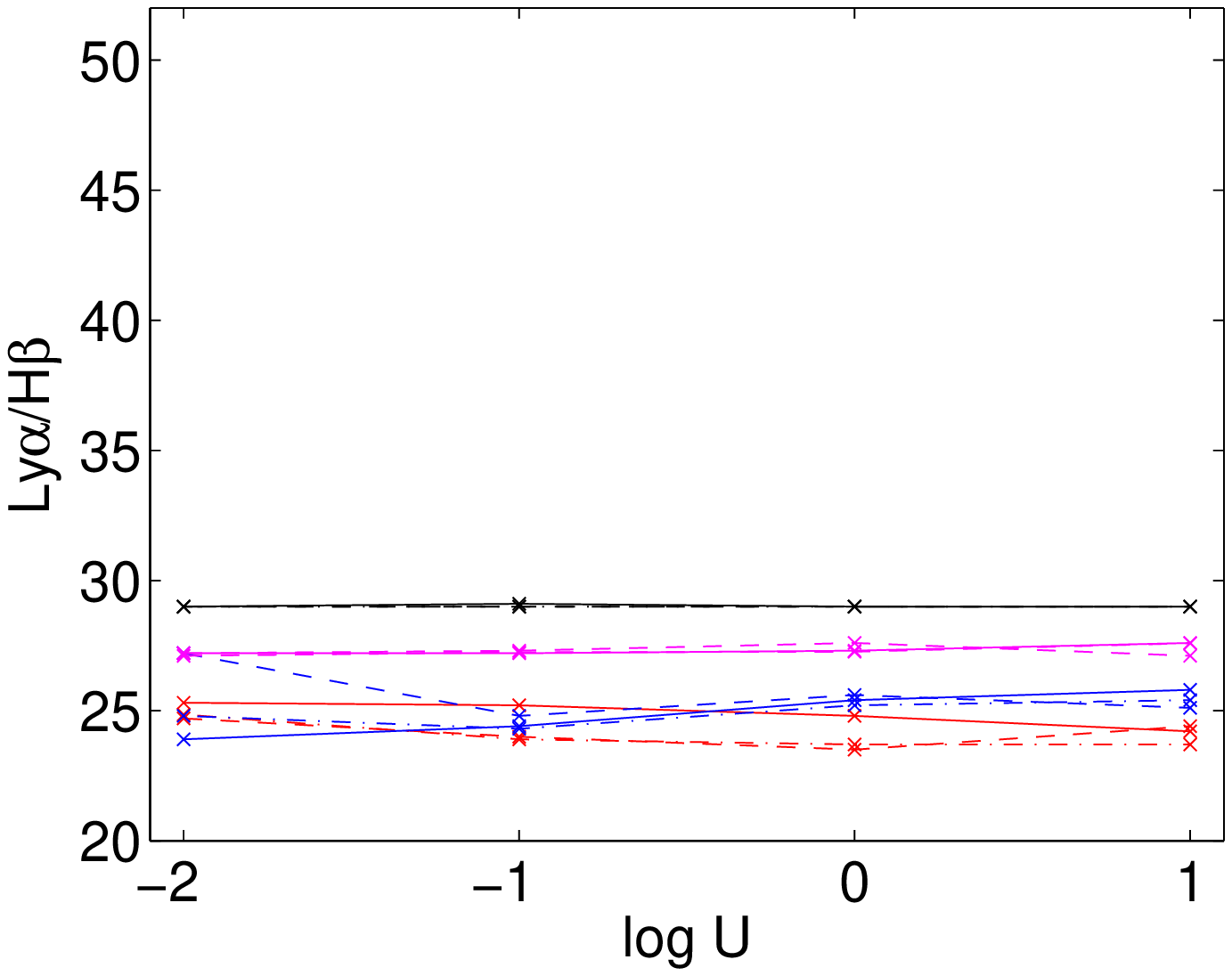}
\includegraphics[width=6.0cm]{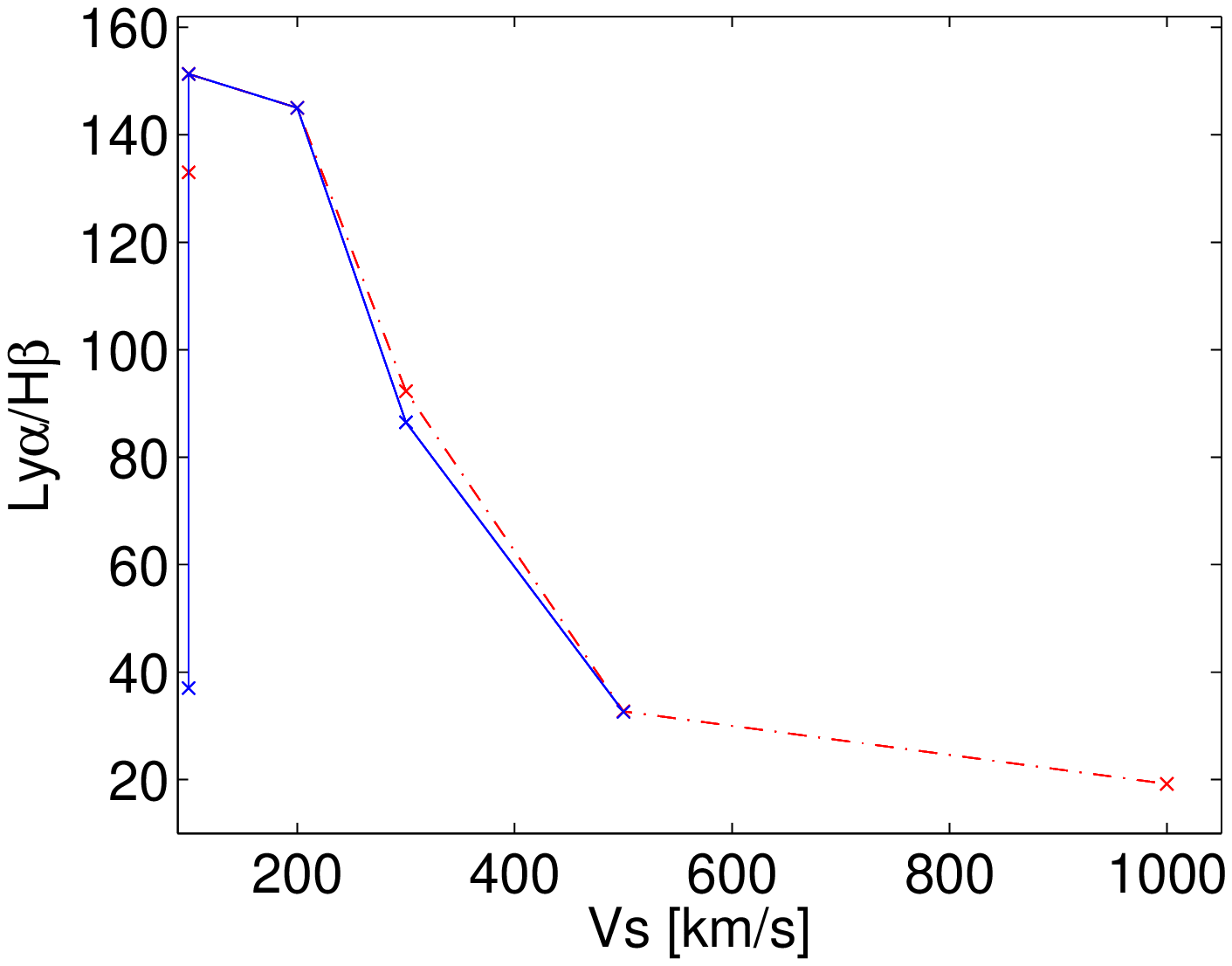}
\caption{Left: the \Ly/\Hb ratio calculated by the grid of models for the NLR of AGN (symbols are given in Table 1);
centre : for SB (symbols in Table 1) right : for shock dominated spectra (red line : D=10$^{17}$ cm; 
blue line : D=10$^{19}$ cm)
}

\end{figure*}

\begin{table}
\caption{The symbols in Fig. 1}
\tiny{
\begin{tabular}{lccccccc} \hline  \hline
\           &     \Vs  & \n0   & log\Ts  &   log$D$  &      symbols        \\
\           &   \kms   & \cm3  & [10$^4$ K] &   [cm]    &                \\ \hline
\   AGN     &     100  &   100 &      -    &       17   &      black dotted         \\
\           &     100  &   100 &      -    &       19   &      black dash-dotted\\
\           &     200  &   200 &      -    &       18   &      blue dotted\\
\           &     200  &   200 &      -    &       19   &      blue dash-dot\\
\           &     100  &   300 &      -    &       17   &      magenta dotted\\
\           &     100  &   300 &      -    &       19   &      magenta dash-dot\\
\           &     300  &   300 &      -    &       17   &      green dotted\\
\           &     300  &   300 &      -    &       19   &      green dash-dot\\
\           &     500  &   300 &      -    &       18   &      red  dotted   \\
\           &     500  &   300 &      -    &       19   &      red  dash-dot\\
\   SB      &     100  &   100 &    4.0    &       19   &      red solid \\
\           &     100  &   100 &   4.7     &       19   &      red dash-dot\\
\           &     100  &   100 &   5.0     &       19   &      red dash  \\
\           &     200  &   200 &    4.0    &       19   &     blue solid \\
\           &     200  &   200 &   4.7     &       19   &     blue dash-dot\\
\           &     200  &   200 &   5.0     &       19   &     blue dash  \\
\           &     300  &   300 &    4.0    &       19   &   magenta solid \\
\           &     300  &   300 &   4.7     &       19   & magenta  dash-dot\\
\           &     300  &   300 &   5.0     &       19   & magenta  dash  \\
\           &     500  &   300 &    4.0    &       19   &    black solid \\
\           &     500  &   300 &   4.7     &       19   &    black dash-dot\\
\           &     500  &   300 &   5.0     &       19   &    black dash  \\ \hline

\end{tabular}}
\label{tab1}
\end{table}

We  calculate for each  galaxy the gas
density, temperature, shock velocity, etc., as well as  the
element abundances and  the photoionizing source.
The  presence of an AGN or/and of a SB, 
collision heating and ionization of the ISM gas by shocks,   the element abundances and dust survival
throughout  the clouds are  important issues  for  middle and high redshifts  galaxies which are  
tracers of star formation and evolution.
The calculation models are  briefly described in Sect. 2.
Modelling  results of the galaxy spectra are presented in Sect. 3. 
Results  and concluding remarks  are discussed in Sect. 4.

\begin{table*}
\centering
\caption{Modelling  the  spectra observed by Humphrey et al (2008) }
\tiny{
\begin{tabular}{llccccccccccccc} \hline  \hline
\              & 0211-122  $^1$   & FWHM$^2$ & MH1  & MH2     &MH3    & av          &0406-244 $^1$  &MH4     & MH5  &  av        \\ \hline         
\     z        & 2.340            &         &      &         &      &               & 2.440         &        &      &      \\           
\   SFR(\Moy)  & 24.85            &         &      &         &      &               & 911.          &        &      &       \\                    
\ \Ly 1216     & 0.21$\pm$0.01   & 1000  & 9.7 & 13.3    & 9.9    & 10.0            & 7.0$\pm$0.4       & 25.6 &  21.2&25.5      \\ 
\  NV 1240     & 0.35$\pm$0.03   &       & 1. &  3.87  & 0.005 & 0.2               &  $<$0.2           & 0.005& 10.9 &0.25        \\  
\ SiIV+OIV 1403& 0.074$\pm$0.008 &       &0.33&  3.33   & 0.01   & 0.15       & 0.3$\pm$0.1      & 0.04 &10.0  &0.27         \\ 
\ NIV]1485     & 0.024$\pm$0.005 &       &0.07 & 0.83  & 0.007  & 0.04         &$<$0.1            & 0.02 & 2.17 &0.069         \\
\ CIV 1549+    & 0.65$\pm$0.02   &       &1.  & 7.67  & 0.15   & 0.5           & 0.9$\pm$0.1      & 0.31 &20.7  &0.772        \\  
\ HeII 1640    & 0.36$\pm$0.01   & 600   &0.83 & 0.67 & 0.4    & 0.42          & 1.2$\pm$0.1      & 1.44 &5.27  &1.53          \\
\ OIII]1667    & 0.028$\pm$0.003 &       & 0.13& 1.23  & 0.04   & 0.1          & 0.2$\pm$0.1      & 0.09 &3.6   &0.17         \\ 
\ CIII] 1909   & 0.17$\pm$0.03   &       & 0.33& 3.67 & 0.46   & 0.5           & 0.9$\pm$0.1      & 0.8  &9.9   &1.0          \\ 
\ CII] 2326    & -               &       & 0.2 & 1.57   &0.1    & 0.17         &0.6$\pm$0.1       & 0.46&3.85   &0.54         \\ 
\ [NeIV]2423   & 0.11$\pm$0.01   &       & 0.07& 0.7   &0.03   &  0.06         &1.0$\pm$0.3       & 0.09 &8.0   &0.3          \\ 
\ [NeV] 3426   &$<$0.3           &       & 0.07& 0.7  &0.02   & 0.05           &0.5$\pm$0.1       & 0.05 &1.8   &0.1          \\ 
\ [OII]3728+   & 0.39$\pm$0.08   & 730   & 0.17& 1.33  & 0.45  & 0.48          &2.6$\pm$0.2       & 2.65 &3.7   & 2.67        \\ 
\ [NeIII]3869  & 0.31$\pm$0.08   & 390   & 0.13& 0.67  & 0.34  & 0.36         &1.0$\$\pm$0.1       & 1.13 &2.0   & 1.15      \\ 
\ [OIII]4363    & $<$0.4        &        &0.02  & 0.2  & 0.032 & 0.04         &    -               &  -   &  -   &   -    \\ 
\  HeII 4686   & $<$0.2          &       &0.07 & 0.07   & 0.03  & 0.036        &0.21$\pm$0.06      & 0.2  &0.4   & 0.21        \\
\ \Hb quiescent  & $<$0.5        &       &0.33& 0.33   & 0.33 & 0.33          & 1.$\pm$0.7$^4$     & 1.   & 1.   & 1.           \\
\ [OIII] 5007+ quiescent& 2.81$\pm$0.1  &$<$ 500&0.4 & 2.67&4.3 & 4.2          &12.6$\pm$0.1     & 13.  &10.0  &12.9           \\
\ [OIII] 5007+ perturbed&1.61$\pm$0.06  &1300   & -  & - &  -  &      -       &   -             & -    & -    & -          \\
\ [OI]         & $<$0.2          &       & 0.10& 0.67  & 0.09  & 0.11         & -                & 1.3  & 0.58 &1.28           \\
\ \Ha quiescent & 0.77$\pm$0.15$^3$ &$<$ 500&1  & 1    & 1     & 1.0           &  3.3            & 2.9  & 3.1  &2.9            \\
\ \Ha perturbed & 0.23$\pm$0.08  &   -   &  - &    -    &   -   &       -     &       -         &    - & -    &  -        \\
\ [NII] 6585+  & $<$0.2          &  -    &0.43 & 1.33  & 0.58  & 0.6        & -               & -    &  -   & -             \\
\ [SII] 6725+  & $<$0.2          &  -    &0.013& 0.13    & 0.05  & 0.055     & -               &   -  & -    & -             \\
\  \Vs (\kms)  & $<$0.2          &  -  &   1000 & 500     & 300   &   -      &  -              & 100  & 700  &-               \\
\ \n0  (\cm3)  &     -           &  -    & 400  & 300     & 300   &    -     &  -              & 100  & 300  &-               \\
\ F $^5$    &      -             & -     &-    &  -      & 20     &   -      &  -              & 1.3  & -    & -   \\           
\ $D$ $^6$     &     -           &  -    & 1  &  1    & 0.01      &   -      &  -              & 1    & 0.028&-      \\         
\ C/H $^7$     &     -           &  -    & 3.3& 3.3     &3.3?     &   -      &  -              & 3.3  & 3.3  & -       \\       
\ N/H $^7$     &     -           &  -    & 1  &  1      &1        &   -       &  -              & 1.   & 1.   & -        \\      
\ O/H $^7$     &     -           &  -    & 6.6&  6.6    & 6.6     &   -       &  -              & 6.6  & 6.6  & -        \\      
\ S/H $^7$     & -               &  -    &  - &   -     &  -      &   -      &   -              & 0.02 & 0.02  &-         \\ 
\ \Hb calc$^8$ &     -           &  -    & 0.1  &0.037 &  1.1     &   -     &  -              & 0.05 & 0.006& -  \\             
\ w            &    -            &  -    & 0.014  &  0.58   &0.40 &   -      &  -              & 0.833& 0.167  & -    \\  \hline   
\end{tabular}}

$^1$ in 10$^{-16}$ \erg;
$^2$ in \kms;
$^3$ flux of narrow \Ha is 6.4$\pm$0.5 10$^{-16}$ \erg; 
$^4$ flux of \Hb is 3.3$\pm$0.2 10$^{-17}$; 
$^5$ in 10$^{10}$ photons cm$^{-2}$ s$^{-1}$ eV$^{-1}$ at the Lyman limit;  
$^6$ in 10$^{19}$ cm;
$^7$ in 10$^{-4}$ units;
$^8$ in \erg

\label{tab2}
\end{table*}

\begin{table*}
\centering
\caption{Modelling  the  spectra by Humphrey et al (2008) }
\tiny{
\begin{tabular}{lccccccccccccccccccc} \hline  \hline
\ flux$^1$         &0529-549 &MH6  &  MH7 & 0828 & MH8  & MH9 &av    & 1138-262& MH10 & MH11 & 4C+23.56 &MH12 &MH13   \\    \hline
\     z            &  2.575  &     &      &  2.572    &      & &        & 2.156   &      &      & 2.479    &     &        \\
\    SFR(\Moy)     &  150    &     &      & 86.        &     & &         &  56.    &      &      & 163.     &      &         \\
\ \Ly 1216              
 & -  & 27.0   & 21.2 & 21.7 & 45.  & 32.5  &33.8& -       & 29.  &27.6  & 2.17   &  47.5 & 30.4    \\
\  NV 1240               
 &   -   & 0.07    & 10.9 & 0.9  & 24.  & 0.02  &2.6& -     & 0.024&0.017 & 0.31  &  0.03+ & 0.06+     \\
\ SiIV+OIV 1403         
 & -      & 0.09   & 10.  & 0.78 & 2.3  & 0.05  &0.3 & -       & 0.02 &0.013 & 0.05  &  0.19  & 0.13    \\
\ NIV]1485             
 & -      & 0.02   & 2.17 & 0.18 & 1.3  & 0.02  &0.16& -       & 0.002&0.002 & 0.06  &  0.08  & 0.02    \\
\ CIV 1549+              
 & -      & 0.2    & 20.7 & 4.5  & 30.3 & 0.2   &3.5& -       & 0.054&0.01  & 0.51  &  1.63  & 0.33    \\
\ HeII 1640            
 & -      & 0.83   & 5.27 & 2.57 & 5.8  & 1.    &1.5& -       & 0.6  &0.011 & 0.43  &  3.7   & 1.32   \\
\ OIII]1667              
 & -      & 0.065  & 3.6  & 0.46 & 3.15 &0.1    &0.43& -       & 0.04 &0.009 & 0.11  &  0.37  & 0.11    \\
\ CIII] 1909              
 & -      & 0.55   & 9.9  & 1.4  & 4.2  &1.1    &1.43& -       & 0.46 &0.1   & 0.28  &  0.7   & 1.   \\
\ CII] 2326             
 & -      & 0.56   & 3.85 & 0.48 & 3.4  &0.9    &1.17& -       & 0.35 &0.1   & 0.22  &  0.19  & 0.85   \\
\ [NeIV]2423              
 & -      & 0.04   & 8.   & 1.2  & 2.   &0.05   &0.26& -       & 0.02 &0.0027& 0.37  &  0.26  & 0.06    \\
\ [NeV] 3426            
 & $<$2.1 &0.018   & 1.8  & 0.76 & 8.85 &0.012  &0.98& -       & 0.01 &0.0028& 0.74  &  0.2   & 0.    \\
\ [OII]3728+           
 & 3.2    & 3.3    & 3.7  & 1.75 & 2.1  &3.7&3.5    & 1.44    & 1.35 &1.2   & 2.57  &  2.7   & 3.4   \\
\ [NeIII]3869        
 & $<$ 2.1 & 1.0   & 2.   & -    & 1.6  &1.48&1.5   & 0.82    & 0.68 &0.29  & 0.74  &  1.36  & 1.5    \\
\ [OIII] 4363
 & $<$1    & 0.04. & 0.63 &   -  &  -   &   - &-  &    -    &  -   &  -   &   -   &   -    &  -     \\
\  HeII 4686           
 & $<$1    & 0.12  & 0.4  & 0.29 & 0.44 &0.15&0.18   & $<$0.88 & 0.09 &0.002 &$<$0.86&  0.53  & 0.2    \\
\ \Hb                   
 &  1.     & 1.    & 1    & 1    & 1    &1    &1  & 1       & 1    & 1    & 1     &  1     & 1    \\
\ [OIII] 5007+            
 & 10.7   & 9.0    & 10   & 9.53 & 8.   & 11.8 &11.4 & 6.      & 6.3  &6.16  & 13.05 &  12.5  & 13.5   \\
\ [OI]   6300+   
 & 0.58   & 1.1    &  -   &   -  &  -   &   -   &-& $<$0.59 &0.5   &0.002 & $<$0.28&0.7    & 1    \\
\ \Ha                     
 & 5.2    &3.0    & 3.1  & 3  & 3      & 3     & 3 &3      & 3    &3     & 3     &3       & 3   \\
\ [NII] 6585+               
 & 2.6     &2.76   &  0.  & -   &  -    &  -    &-& 2.35    & 1.9  &0.8   & 0.8   &1.0     & 1.86  \\
\ [SII] 6725               
 & 1.58    &0.6    & 0.   & -   &  -    &  -    &-& 0.85    & 0.63 &0.1   & 0.86  &0.8     & 0.7  \\
\  \Vs (\kms)                
 & -       &180    & 700  & -  & 700    & 1200  &-& -       & 300  &300   & -     &200     & 200  \\
\ \n0  (\cm3)                 
 & -       &100    & 300  & -  & 400    & 400   &-& -       & 300  &300   & -     &300     & 300  \\
\  \Ts $^2$                   
 &  -      & -     &  -   &  - &  -     &  -    &-&  -      &  -    &  4  & -     & 25.    & -  \\
\ $U$                        
 &  -      &  -    &  -   &  - &  -     &  -    &-&  -      &  -    & 0.8 & -     & 0.16   & - \\
\ F $^3$                   
 & -       &1.0    & -    & -  & -      & 6     &-&  -      & 8     & -   & -    & -      & 4   \\
\ $D$ $^4$                  
 & -       &1      & 0.028& -  & 0.019  & 0.3   &-&  -      & 0.01 &0.01  & -    & 1.     & 0.01   \\
\ C/H $^5$                 
 & -       &3.3    & 3.3  & -  & 3.3    & 3.3   &-&  -      & 3.3  &3.3   & -    & 0.8    & 0.8   \\
\ N/H $^5$                  
 & -       &1      & 1    & -  & 1      & 1     &-&  -      & 1.   & 1    & -    & 0.5    & 0.5   \\
\ O/H $^5$                 
 & -       & 6.6   & 6.6  & -  & 6.6    & 6.6   &-&  -      & 6.6  & 6.6  & -    & 6.6    & 6.6   \\
\ S/H $^5$                
 &  -      & -     &  -   & -  & -      & -     &-&   -     & 0.3  & 0.3  & -    & 0.3    & 0.3   \\
\ \Hb calc $^6$               
 & -       &0.06   & 0.0058 & -& 0.0089 & 0.29  &-&  -      & 0.62 & 1.6  & -    & 0.59   & 0.14    \\ 
\  w                     
 & -       & -     &   -    & -&  0.8    & 0.2  & -&  -     &   -  &  -   &  -   &  -     &  -    \\ \hline
\end{tabular}}

$^1$ in 10$^{-16}$ \erg; $^2$ in 10$^4$ K; $^3$ in 10$^{10}$ photons cm$^{-2}$ s$^{-1}$ eV$^{-1}$ at the Lyman limit; $^4$ in 10$^{19}$ cm;
$^5$ in 10$^{-4}$ units; $^6$ in \erg.

\label{tab3}
\end{table*}

\section{The calculation code}

The line and continuum spectra emitted by  gas  and dust
are calculated by the
code {\sc suma}\footnote{http://wise-obs.tau.ac.il/$\sim$marcel/suma/index.htm}.
The code simulates the physical conditions in an emitting gaseous cloud under the coupled effect of 
photoionization from an external radiation source and shocks. The line and continuum emission 
from the gas are calculated consistently with dust-reprocessed radiation in a plane-parallel geometry.

\subsection{Input parameters}

 To calculate  line  and continuum fluxes  emitted from a gas nebula,  the  temperature, density
 and   element abundances should be known. 
 In a shock dominated hydrodynamical regime
the  density profile  throughout  the nebula is shaped by the shock,
while   radiative and  collisional heating and ionization of the gas  are due to radiation 
from  the external source  (AGN, SB) and by the shock, respectively.
The  input parameters which characterise the shock are roughly suggested by the data, e.g.
the  shock velocity \Vs by the  FWHM of line profile and  the pre-shock density \n0 by  the characteristic line ratios 
and by the pre-shock magnetic field \B0. 
We   adopt   \B0  = 10$^{-4}$ G, which is suitable to the NLR of AGN (Beck 2011).

The ionizing radiation from an external source is characterized by its spectrum  and by 
the flux intensity. The flux is calculated at 440 energies, from a few eV to keV. 
If the photoionization source is an active nucleus,  the input parameter  that  refers 
to the radiation field  is the power-law
flux  from the active center $F$  in number of photons cm$^{-2}$ s$^{-1}$ eV$^{-1}$ at the Lyman limit
with  spectral indices  $\alpha_{UV}$=-1.5
and $\alpha_X$=-0.7.
  It was found by modelling the spectra of many different AGNs that these indices  were the most
suitable, in general. (see, e.g.  Contini \& Aldrovandi 1983, 
Aldrovandi \& Contini 1984,   Rodr\'{i}guez-Ardila et al 2005, and references therein).
In particular,  for all the models presented in the following, we use $\alpha_{UV}$ = -1.5, 
recalling that the shocked zone 
also contributes to the emission-line intensities, so that our results are less dependent on the shape 
of the ionizing radiation. 
The power-law in the X-ray domain  was found flatter by the observations of local galaxies
 (e.g. Crenshaw et al 2002, Turner et al 2001).
 $F$  is combined with the ionization parameter $U$ by :
$U$= ($F$/(n c ($\alpha_{UV}$ -1)) (($E_H)^{-\alpha_{UV} +1}$ - $(E_C)^{-\alpha_{UV} +1}$)
(Contini \& Aldrovandi, 1983), where
$E_H$ is H ionization potential  and $E_C$ is the high energy cutoff,
$n$ the density, $\alpha_{UV}$ the spectral index, 
and c the speed of light.
If  stars with a colour temperature \Ts are the photoionization source,
the number of ionizing photons cm$^{-2}$ s$^{-1}$ produced by the hot 
source is $N$=$\int_{\nu_0}^{\infty}$$B_{\nu}$/h$\nu$ d$\nu$, 
where $\nu_0$ = 3.29  10$^{15}$ Hz and B$_{\nu}$  is the Planck function. 
The flux from the star is combined with $U$ and n by $N$ (r/R)$^2$=$U$nc, where r is the radius of the hot source,
 R is the radius of the nebula (in terms of the distance from the stars). 
Therefore, \Ts  and $U$ compensate each other, but only in a qualitative way, because the black body radiation
(which depends on \Ts)  determines 
the frequency distribution of the primary flux, while $U$ represents the number of photons per number of electrons 
reaching the nebula. The choice of \Ts and $U$   is made by the fit of the line ratios.

In addition to the radiation from the primary source, the effect of the diffuse radiation created by 
the hot gas  is also calculated, using 240 energies  for the spectrum.
The geometrical thickness of the emitting nebula $D$  determines whether the model is
radiation-bounded or matter-bounded.
The dust-to-gas ratio ($d/g$) and the  abundances of He, C, N, O, Ne, Mg, Si, S, A, Fe relative to H,
are also accounted for. 
 We adopt an initial grain radius \agr=1\mum.

\subsection{Calculation process}

The code accounts for the direction of the cloud motion relative to
  the external photoionizing source.  A parameter switches between  inflow (the  
radiation flux from the source
reaches the  shock front edge of the cloud)  and outflow (the flux reaches the
edge opposite to the shock front).
The calculations start at the shock front 
 where the gas is compressed and thermalized adiabatically, 
reaching the maximum temperature in the immediate post-shock region (T$\sim 1.5 \times 10^5$ (\Vs/100 \kms )$^2$). 
T decreases downstream  by the cooling rate and the gas recombines. 
The downstream region is cut  into a maximum of 300 plane-parallel slabs with different geometrical 
widths calculated automatically, in order to account for the temperature gradient (Contini 2009 and references therein).
In each slab, compression ($n/n_0$) is calculated by the Rankine--Hugoniot equations for the conservation of mass, 
momentum and energy throughout the shock front (Cox 1972). Compression  ranges between 4 
(the adiabatic jump) and $>$ 100, depending on \Vs and \B0. The stronger the magnetic field, the lower the 
compression downstream, while a higher shock velocity corresponds to a higher compression.
The cooling rate is calculated in each slab by free-free (bremsstrahlung), free-bound and line emission. 
Therefore,   most emission lines must be calculated  in each slab even if only a few ones
are observed because they contribute to the temperature slope downstream.

In pure photoionization models, the density $n$ is constant throughout the nebula.
In models accounting for the shocks, both the electron temperature  $T_{\rm e}$  and  density \ne 
show a characteristic profile throughout each cloud. 
The density reaches its upper limit  at a certain distance from shock-front and  remains nearly constant,
while  \ne decreases following recombination.

The primary and secondary radiation spectra, calculated by radiation transfer, change throughout the downstream slabs, 
each of them contributing to the optical depth. 
In each slab of gas the fractional abundance of the ions of each chemical element is obtained by solving 
the ionization equations which account for   
photoionization (by the primary and diffuse secondary radiations and collisional ionization) and for  recombination 
(radiative, dielectronic), as well as for charge transfer effects, etc. 
The ionization equations are coupled to the energy equation when collision processes dominate (Cox 1972) and 
to the thermal balance if radiative processes dominate (Williams 1967). The latter balances the heating of the gas due 
to the primary and diffuse radiations reaching the slab with the cooling due to  
 line emission, dust collisional ionization and thermal bremsstrahlung. 
The  line intensity contributions
from all the slabs are integrated throughout the cloud.
In particular, the absolute line fluxes referring to the ionization level i of element K are calculated by the 
term $n_K$(i) which represents the density of the ion X(i). We consider that $n_K$(i)=X(i)[K/H]$n_H$, where X(i) is 
the fractional abundance of the ion i calculated by the ionization equations, [K/H] is the relative 
abundance of the element K to H and $n_H$ is the density of H (by number \cm3). In models including shock, 
$n_H$ is calculated by the compression equation  in each slab downstream. 
So the element abundances  relative to H appear as  input parameters.
To obtain   the N/H relative abundance for each galaxy, we consider  the
charge exchange reaction  N$^+$+H $\rightleftharpoons$ N+H$^+$
(Steigman et al. 1971). Charge exchange reactions occur between ions with similar ionization potential
(I(H$^+$)=13.54 eV, I(N$^+$)=14.49 eV  and I(O$^+$)=13.56 eV). It was found that N  ionization equilibrium in the ISM
is strongly affected by  charge exchange.  This  process
as well as  O$^+$+H $\rightleftharpoons$ O+H$^+$  (Field \& Steigman 1971) are included in  the SUMA code.
The N$^+$/N ion fractional abundance follows the behaviour of O$^+$/O so, comparing the
[NII]/\Hb   and the [OII]/\Hb line ratios with the data,
the N/H relative abundances can be easily determined (see Contini et al. 2012).

Dust grains are coupled to the gas across the shock front by the magnetic field. 
 They  are heated radiatively by  photoionization  and collisionally by the gas.
up to the evaporation temperature ($T_{\rm dust}$ $\geq$ 1500 K). 
The distribution of the grain radii downstream is determined 
by sputtering (Dwek 1981), which depends on the shock velocity and on the gas density. 
Throughout shock fronts and downstream, 
the grains might be completely destroyed by sputtering.
The dust-to gas ratio $d/g$  affects the mutual heating and cooling of gas and dust.

The calculations 
proceed until the gas  cools down to a temperature
below 10$^3$ K (the model is radiation bounded) or the calculations are interrupted when all the lines reproduce
the observed line ratios (the model is matter bounded).
In  case that photoionization and shocks act on opposite edges, i.e. when the cloud propagates outwards from the 
radiation source, the calculations
require some iterations, until the results converge.
In this case  the cloud geometrical thickness plays an important role.  
Actually, if the cloud is
very thin, the cool gas region may disappear  leading to low or  negligible low ionization level lines. 

Summarizing, the code starts in the first gas slab  adopting  
 the input parameters  given by the model.
 Then, it calculates the density, the fractional abundances of the ions from each level for each element,  
free-free,  free-bound  and line emission fluxes. It calculates \Te by thermal balancing or the 
enthalpy equation, and  the optical depth of the slab  in order to obtain the primary and 
secondary fluxes by radiation transfer for the next slab. Finally,  the parameters  calculated in slab i 
are adopted  as initial conditions 
for slab i+1. Integrating the  line intensities  from each slab, 
 the absolute fluxes of  the lines and  of bremsstrahlung  are obtained at the nebula. 
The line ratios to a certain line (generally \Hb for the optical-UV spectrum) are then calculated
and  compared  with the observed data, in order to avoid problems of distances,  absorption, etc.
 The number of the  lines calculated by the code (more than 300) does not depend on the number of
the observed lines nor on the number of  input parameters,
but  on the elements composing the gas.

\subsection{Selection of the models. The grids}

The  physical parameters are combined throughout the calculation of forbidden and permitted lines
 emitted from a shocked nebula.
The ranges of the physical conditions in the  gas are deduced, as a first guess, from the observed
line ratios 
 because they are more constraining than the continuum SED.
We make use of 
the grids presented by Contini \& Viegas (2001a,b) for AGN (GRID1) and SB (GRID2), respectively
for a first estimate of the gas physical conditions. The grids were
calculated by a gradual increase of the  input parameters   adapted  from  the  observed issues.
Grid results give a  rough approximation to the observed spectra. 
Diagnostic diagrams for \Ly/\Hb calculated by the code SUMA 
for ejected clouds (Fig. 1) show that
radiation dominated models (+shocks) lead to \Ly/\Hb between 20 and 30, with some higher values (up to 50)
for low velocity, high density extended clouds. For shock dominated clouds \Ly/\Hb ranges between  $\sim$150 and $\sim$15.

The  parameters are then 
refined by the detailed modelling of  the spectrum.
Specific grids of  models are calculated for each galaxy
modifying  the parameters on small scales  in order to  reproduce as close as possible all  the observed  line ratios
in each spectrum.
We generally
consider that the observed spectrum is "satisfactorily" fitted by a model when the strongest lines are
reproduced by the calculation within  20\% and the weak ones within 50\%.
  "Satisfactorily"   means that
the  approximation can be accepted 
1) in the context of limits  set by the author community,
2)  considering the ensemble of  all the line fit,
3) within the limits of  the observational error,
4) if it can be improved only adopting  unsuitable  gas conditions, and
5)  because, nevertheless,  adding  some information. 

 The final gap between  observed and  calculated line ratios  is  due to the  observational errors  both random
and systematic, as well as to the uncertainties of the atomic parameters  adopted by the code,
such as recombination coefficients, collision strengths etc., which are continuously updated,
and to the choice of the model itself.

The observed spectra provide averages throughout the whole galaxy, particularly at high redshifts.
When  the  observed line ensemble  is not satisfactorily reproduced by one model,
the contribution  of  other regions  within the galaxy at  different physical conditions must be 
accounted for, leading to pluri-cloud models. 
The different results  are summed up  by  relative weights
 (see, e.g. Rodriguez-Ardila et al. 2005)
 when they compensate each other (e.g. in Table 2). When they show similar results they are left as
alternative proposals (e.g. Table 3).

The models selected   by the fit of the line spectrum  are cross-checked by  fitting the continuum SED.
In the  UV range the  bremsstrahlung from the  nebula is   blended with  black body emission 
from  the  star population background. 
The maximum frequency of the bremsstrahlung peak in the UV - X-ray domain
depends  on the shock velocity. 
In the IR range  dust reprocessed radiation is generally seen (Contini, Viegas \& Prieto 2004).  
 In the radio range synchrotron radiation by the Fermi mechanism at the
shock front is easily recognized by the slope of the SED.

The set of the input parameters which leads to the  best fit of the observed
line ratios  determines the physical and chemical  properties of the emitting gas. They are considered as
the "results"  of modelling.

\begin{table*}
\centering
\caption{Modelling UV spectra by Dawson et al (2003), Stern et al (2002) and Erb et al (2010)} 
\tiny{
\begin{tabular}{lccccccccccccccccc} \hline  \hline
\              &    HDF$^1$ &          FWHM$^2$ &  MDS   & CX0$^3$        & FWHM$^2$ &  MSS         &MSR    & BX418$^4$ &FWHM  & ME1-0&ME2-0& ME3 &MESB  \\ \hline
\    z         &    2.011       &               &        & 3.288          &         &                &      & 2.3   &  &      &     &     &        \\
\ SFR (\Moy)   &    4.73        &               &             & 18.75          &         &          &       & 17.5  &  &      &     &     &       \\
\ OVI 1035     &      -         &         -     & 44.4        &  1.5:          &2640     &  12.8     &0.25  & -     &-  &   -  &  -  & -   &  -    \\
\ OV  1215     &      -         &          -     & 2.4        & -              &  -      &  -        &-     & -     &-  &   -  &-    &-    &  -    \\
\ \Ly 1216     &      1.68      &        1270    &  1.5       & 18.9           &1520     &  29.9     &35.   & 11.27 &840  & 35.  &34.5 & 29.6& 26.6  \\
\  NV 1240     &      2.3       &        2100      & 2.2      & 0.6:            &1820     &  0.65     &0.16 & -     &-  & -    &-    &-    &  -   \\
\ SiIV+OIV 1403&        -       &         -      &   -       &0.4               &1320     &  0.3+0.86 &0.22 & -     &-  & -    &-    &-    &  -   \\
\ CIV 1549+    &      2.47      &         1300   &      2.2  &3.5               &1350     &  3.5      &3.4  & -     &-  & -    &-    &-    &  -   \\
\ HeII 1640    &      0.45      &        1400    &   1.0     &1.7               &940      &  2.       &3.0  & 0.31  &612  & 0.3  &0.31 &0.89 &  0.003\\
\ OIII]1667    &      -         &           -    &    -      & 0.9             &1290     &  0.3      & 0.4  & 0.23  &235&0.033 & 0.06&0.14 &  0.02 \\
\ CIII] 1909   &      0.19      &        900     & 0.3       & 2.1             &1090     &  2.       &1.9   & 0.54  &225  & 0.16 & 0.27& 0.88&  0.2  \\
\ NeIV] 2424   &      0.3       &        1470    & 0.20      &  -              &-        &   -       &   -  & -     &-  & -    & -   & -   &  -    \\
\ [OII] 3727   &       -        &        -       &           &                 &         &   -       &      & $<$0.73&-225& 0.045& 0.09&0.65 &  0.66  \\
\ \Hb    4861  &       -        &          -     & 0.1       & 1.2             &170      & 1         &1     & 1      &$<$102 & 1    & 1   & 1   &  1    \\
\ [OIII] 5007+ &        -       &          -     & 0.7       &18.1             &300-430  &2.7        & 17.1 & 8.5    &140 & 0.2  & 0.49&8.   &  8.88 \\
\ [NII] 6584+6548 &      0.96   &        380     &    0.02   &  -              &-         &   -      & 2.74 & $<$0.12&- & 0.28 & 0.48& 0.2 & 0.22  \\
\ \Ha n        &     0.7        &        240     & -         &  -              &-         &  -       & 3.   & 3.08   &155&3.   & 2.9 &3.3  & 3.     \\
\ \Ha b        &     2.3        &        2500    &  0.278       &  -             &-         & 2.7      &  - &  -     &- & -    & -   &-    &   -   \\
\  \Vs (\kms)  &      -         &         -      & 1300       & -               &-       &1500       &200   & -      &- & 840  & 600 & 260 &  260   \\
\ \n0  (\cm3)  &    -           &         -     &300         &  -               &-      &110        &350    & -      &- & 3000 & 3000& 200 &  200   \\
\ F $^5$       &   -            &         -      &0          & -               & -      &-          &100    & -      &- & 50   & 50  & 30  &  -      \\
\ \Ts$^6$      &   -            &         -      & 2.5       &  -              &        &-          &  -    & -      &- & -    & -    & -   & 6.    \\
\ U            &    -           &         -     & 8          &  -               &       &-          &  -    & -      &- & -    & -    & -    & 0.7   \\
\ $D$ (10$^{16}$cm)  &     -     &          -    &9.0        &   -             &-       &41.5      &100     & -      &- & 3.6  & 2. &  400 & 400   \\
\ C/H $^7$     &     -          &        -      &2.6        &  -               &-      &3.3        &3.3     & -      &- &3.3   & 3.3 & 3.3   & 3.3 \\
\ N/H $^7$     &     -          &        -      &1.         &  -               &-      &0.4        &0.1     & -      &- &0.3   & 1.0 & 0.3   & 0.3 \\
\ O/H $^7$     &    -           &        -      &6.6        &  -               &-      &3.6        &6.6     & -      &- &6.6   & 6.6 & 6.6   & 6.6  \\
\ \Hb calc$^8$ &    -           &        -      &0.0085     &  -             - & -      &0.035      &1.77   & -      &- &10.8  & 4.5 & 1.65  & 0.91 \\ \hline
\end{tabular}}

$^1$ HDFx28 (Dawson et al 2003) in 10$^{-17}$ \erg;
$^2$ in \kms;
$^3$ CXO J084837.9+445352 (Stern et al 2002) in 10$^{-17}$ \erg;
$^4$ BX418 (Erb et al 2010) in 2.6 10$^{-17}$ \erg;
$^5$ in 10$^{10}$ photons cm$^{-2}$ s$^{-1}$ eV$^{-1}$ at the Lyman limit ;
$^6$ in 10$^4$ K
$^7$ in 10$^{-4}$ ;
$^8$ in \erg;

\label{tab4}
\end{table*}

\begin{table*}
\centering
\caption{Modelling  the  spectra from the observations  of Vernet et al  (2001)}
\tiny{
\begin{tabular}{lcccccccccccccc} \hline  \hline
\  Flux$^1$   & 4C+03.24   & MV1   & 0943-242    &MV2 & MV3               & 0731+438&MV4 &MV5  &av   & 4C-00.64     & MV6                 \\ \hline
\   z          & 3.560     &      &2.922           &    &                 & 2.429   &    &      &      &  2.360       &                  \\
\ SFR$^2$    & 192.3     &      & 100.6         &     &                   & 128.6   &    &      &      & 73.27        &   \\
\ OVI     &-          & -    &5.6            &4.3  &9.5                   &-        &-    &-    & -    &  -           & -                  \\
\ OV      &4.8$\pm$0.8&7.3   &  -            &-    & -                     & -       & +   & -   & -    &  -           & -                  \\
\ \Ly    &254.4$\pm$7.3&237.  & 213.8$\pm$1.8&240.& 263.                 &428.1$\pm$2.2&426. &455.&445. & 261.6$\pm$3.4& 233.                 \\
\  NV     &4.8$\pm$0.8& 5.2  & 11.2$\pm$0.6 &4.4 & 14.2                  &7.1$\pm$0.6  &8.0 & 7.7 &7.8  & 12.4$\pm$1.2 & 15.3                \\
\ CII    &4.6$\pm$1.3 & 1.6  &2.4          &0.7 &2.2                     &2.2          &1.4 & 0.7 &0.94 & 1.1          & 1.4                  \\
\ OIV+SiIV&6.1$\pm$2.5&5.0   &8.9          &9.1 & 16.                    &9.7          &13.0& 3.9 &7.0  & 10.4         & 14.                 \\
\ NIV]    &-          &-     &2.4          &4.  & 6.8                    &1.8$\pm$0.3  &3.0 & 2.5 &2.67 & -            &  -                 \\
\ SiII    & -         & -    &1.2          &  - & -                      &0.9          & -  &  -  &-    & -            &   -                \\
\ CIV    &12.5$\pm$2.2&10.5 &46.5$\pm$1.1 &55.2& 40.                    &46.5$\pm$1.0 &17. & 47.3&36.9 & 29.1$\pm$2   &30.07                   \\
\ [NeV]   & -         &-     &-            &0.4 & 0.62                   &-            &-   &  -  &-    & -            & -                    \\
\ HeII  &5.2$\pm$1.3 &6.0    & 47.1$\pm$1.3 &71.8& 22.8                  &30.4$\pm$0.5 &23. & 19.6&20.8 &21.5$\pm$0.5  &  11.2                   \\
\ OIII]  & -         & -    &7.6          &4.  &6.                      &6.4          &1.7 & 2.4 &2.16 &3.6           & 4.                    \\
\ NIII]   & -         & -    &1.4          &2.  & 2.9                   & -           &0.5 & 1.2 &0.96 &-             & -                    \\
\ SiII    & -         & -    &1.7          &-   & -                      &1.4          &-   & -   &-    &-             & -                    \\
\ SiIII]  & -         & -    &5.4          &1.4 & 1.6                   &2.7          &0.1 & 1.1 & 0.8 &-             & -                    \\
\ CIII]   & -         & -    &29.3$\pm$3.1 &24. & 14.                   & 21.2$\pm$0.5&1.4 & 24.8&16.9 & 9.6$\pm$0.7  & 11.7                   \\
\ \Ha     & 30.5      &  -   &25.66        & -  &  -                    & 51.36      & -  &  -  &  -  & 31.4         &  -                 \\
\ \Vs $^3$  & -         & 130  &    -       &100 & 130                   &   -        & 140& 140 & -& -            & 120                    \\
\ \n0 $^4$  & -         &  60  &   -       & 100 & 150                   & -          &150  &130 & -   & -            & 60                     \\
\ F $^5$       & -         & -    &   -       & 2.5 & -                  & -          &-    &6.  & -   & -            & -                     \\
\ \Ts $^6$     & -         & 4.55 &  -       & -   & 6                   & -          &5    & - & -    & -            &4.5                      \\
\ U            & -         & 0.7  &   -       & -   & 0.15               & -          &0.9  & -   & -  & -            &0.7                      \\
\ $D$ $^7$     & -         & 0.013&   -       & 0.01& 0.0008             & -      &0.001  &0.1 & -    & -            &0.01                      \\
\ C/H $^8$     & -         & 3.3  &   -       & 3.3 & 3.3                & -          &3.3  & 3.3& -   & -            &3.3                      \\
\ N/H $^8$     & -         & 1.   &   -       & 1. & 1.                  & -          &1.   &1.  & -   & -            &1.                     \\
\ O/H $^8$     & -         & 6.6  &   -       & 6.6 & 6.6                & -          &6.6  &6.6 & -   & -            &6.6                         \\

\ \Hb $^9$ & -         &0.009 &   -      &0.012 &0.009                 & -         &0.014 &0.13 & -    &-       &2.6e-4\\
   w         &  -         &   -   &-        &  -     &  -                &   -       &0.97  &  0.03 &   -&-&-  \\ \hline
\end{tabular}}

$^1$ in 10$^{-17}$ \erg;
$^2$ in \Moy ;
$^3$ in \kms;
$^4$ in \cm3;
$^5$ in 10$^{10}$ photons cm$^{-2}$ s$^{-1}$ eV$^{-1}$ at the Lyman limit;
$^6$ in 10$^4$ K;
$^7$ in 10$^{17}$ cm;
$^8$ in 10$^{-4}$ units;
$^9$ calculated in  \erg

\label{tab5}
\end{table*}

\section{Modelling single galaxy spectra} 

\begin{table*}
\centering
\caption{Modelling  the  spectrum of the SST24 J1434110+331733 nebula (Dey et al 2005), CDF-S202  (Norman et al 2002) and Lynx arc (Fosbury et al 2003)}
\tiny{
\begin{tabular}{llcccccccccccccc} \hline  \hline
\    flux          & SST24$^1$      &  MDR   &MDOS    &MDSB  &     av       & CDF-S202$^2$ &FWHM$^3$ &  MNR   &MNS  &Lynx$^4$   & MF0  &MFSB \\  \hline
\    z             &   2.656        &        &        &      &             &  3.7          &         &        &     & 3.36      &     &  \\
\ SFR (\Moy)        &   134.6     &       &         &     &             &   136.2        &         &       &     & 177.8     &     &   \\
\ \Ly 1216         & 28.9$\pm$0.15 & -       &   -     & -   &-            &-              &-         &   -   & -   &  -        &  -  & - \\
\ \Ly 1216         & 4.05$\pm$0.01 & -       &   -     &  -  &-            &-              &-         &   -   & -   &  -        &  -  & - \\
\ \Ly 1216         & 3.10$\pm$0.01 & 3.03    &  3.3    &4.9  &  3.39       & 16.4          &$<$1130   &28.    &22.6 & 10.8     &31.8  & 38.    \\
\  NV 1240         & $<$0.05        & 0.017  &  0.03   & 0.1 &      0.03   &  5.9          &$<$1680   &0.     &8.75 & $\leq$0.09&0.3   &0.001\\
\  SiIV 1397        &       -        &  -    &    -    &  -  &     -       &   -           & -        &  -    & -   & $\leq$0.09 &0.09& -\\
\  NIV] 1483+      &       -        &  -    &    -    &  -   &    -        &  -           & -        &  -    & -   &0.42        &0.15  & 0.05 \\
\ CIV 1549+        &0.417$\pm$0.001  & 0.6   &  0.47   & 0.4  &   0.49     & 9.9           &$<$1680    &3.19  &12.2 & 3.65      &3.6  & 0.46 \\
\  HeII 1640       &0.4$\pm$0.004    & 0.82  &  0.39   & 0.3  &   0.45     & 2.8           &$<$680     &0.7   &6.48 &0.11       &0.11 & 0.12 \\
\ OIII] 1661+      &      -          &  -    &   -     &  -   &   -        &    -          &   -      &  -    &  -  & 0.56      &0.48 & 0.09\\
\ NIII] 1750       &      -          &  -    &   -     &  -   &   -        &    -          &   -      &  -    &  -  & 0.18      &0.1  & 0.12\\
\ SiIII] 1883+     &      -          &  -    &   -     &  -   &   -        &    -          &   -      &  -    &  -  & 0.15      &0.14 &0.06\\
\ CIII] 1909       &0.05$\pm$0.01    & 0.0013 & 0.18   & 0.03 &   0.13     &  -            &   -       &12.8   &4.8 & 0.59      &1.4  & 1.0  \\
\ [OII] 3727+      &      -          &  -     &   -    &  -   &  -         &   -           &   -       &  -    &  - &$\leq$0.25 &0.17 & 0.57  \\
\ [NeIII] 3869+    &      -          &  -     &   -    &  -   &  -         &   -           &   -       &  -    &  - &$\leq$0.91 &0.2  &1.16 \\
\ HeII  4686       &       -         &  -     &   -    &  -   &  -         &   -           &   -       &  -    &  - &$\leq$0.22 &0.01 &0.02 \\
\  \Hb             &    -            & 0.1    &  0.1   &  0.1 & 0.1        &   1           &   -       &  1    & 1  & 1         &1.   & 1   \\
\ [OIII] 5007+     &       -         & -      &   -    &  -   &  -         &   -           &   -       &  -    &  - & 10.1      &0.7  & 9.  \\
\       5084       & 0.52$\pm$0.004  &  -     &    -   &  -   & -          &   -           &   -       &   -   &  - &-          &-    & -  \\
\  \Ha  6563       & 0.371           &  0.3   & 0.3    & 0.3  &0.3         &  1.97         &   -       &   3.  &  3.& -         &3.7  &   3.1\\
\  \Vs (\kms)      &  -              & 1700   &1700   &  300  & -          &  -            &  -       &1600    &1600&-          &320  &100    \\
\ \n0  (\cm3)      &  -              & 110    &110    &  150  & -          &  -            &  -       &120     &120 &-          &2000 &300  \\
\ F $^5$           &  -              & 100    & -     & -    &  -          &  -            &  -       & 1      & -  &-          & -   & -   \\
\ \Ts$^6$          &  -              & -      & -     & 2.5   &  -         &  -            &   -      &  -    &  -  &-          &-&1.1   \\
\ U                &  -              & -      & -     &   8   &   -        &  -            &    -     &  -    &  -  &-          &-&0.3   \\
\ $D$ (10$^{17}$ cm) &  -          & 300     & 3.     & 5.     &-          &   -           &  -       & 100   &15   &-          &1 & 4900\\
\ C/H $^7$         &  -              & 3.3    & 3.3   & 3.3   &-           &   -           &  -       & 3.3   & 3.3 &-          &3.3&3.3 \\
\ N/H $^7$         &  -              & 1.     & 0.1   & 1.    &-           &    -          &  -       & 1.    & 1.  &-          &0.2 & 1.0  \\
\ O/H $^7$         &  -              & 6.6    & 6.6   & 6.6   &-           &   -           &  -       & 6.6   & 6.6 &-          &6.6  & 3.6\\
\ \Hb calc$^8$     &  -              & 0.0336 & 0.024 & 5.68   &-          &    -          &  -       & 200. &0.0085&-          &0.15& 8.7   \\
\ w            &  -               &0.137    &0.822 & 0.041 &-  &-&-&-&-&-&-&- \\ \hline
\end{tabular}}

$^1$  10$^{-16}$ \erg  (Dey et al.);
$^2$  10$^{-16}$ \erg; (Norman et al (2002);
$^3$ in \kms ;
$^4$ in 10$^{-17}$ \erg (Fosbury et al 2003);
$^5$ in 10$^{10}$ photons cm$^{-2}$ s$^{-1}$ eV$^{-1}$ at the Lyman limit;
$^6$ in 10$^5$ K
$^7$ in 10$^{-4}$ units;
$^8$ in \erg 

\end{table*}

\subsection{Spectra of the radio galaxies at z$\sim$ 2.5 from the Humphrey et al (2008) survey}

The long-slit NIR spectra obtained by the Infrared Spectrometer And Array Camera (ISAAC) instrument at VLT
and completed with optical lines (by Keck and Subaru)  contain a  collection of lines
in  the wavelength range 1200-7000 \AA. 
The  observed  \Ly/\Hb line ratios bridge between  UV and  optical data.
Humphrey et al, modelling the spectra by pure radiation  and
pure shock  models, concluded that the galaxies are most probably AGN in different physical
conditions,   with roughly solar abundances.
We have   selected   dereddened (\Ha/\Hb $\sim$ 3)  spectra of  galaxies  containing  the  \Hb flux
in order to  compare the \Ly/\Hb line ratios  (calculated at the nebula) with the observation data.  
 We wish to investigate whether an  SB contributes
to  some of the observed line ratios.

We consider  all the
line ratios in each spectrum (Tables 2 and 3), trying  to reproduce them by  models
which account for the coupled effect of photoionization and shocks.  
The models  are described in the bottom of the tables. 
Models  corresponding to $F$=0  are shock dominated,
i.e. the photoionising flux from the source   outside the emitting nebula is hidden or absent.  
Similar shock dominated models are used  to model  supernova remnant spectra.
  The FWHM of the line profiles   gives a hint about the  shock velocity. 
Different FWHM of the  lines in the UV and in the optical range indicate that
the spectra emitted  from different  clouds within the galaxy should be considered.
So pluri-cloud models are adopted.
The weighted sum of the line ratios calculated by  
 different theoretical models for each spectrum (e.g. col. 7 and 11 in Table 2) is compared with the observed 
line ratios.
The weights adopted  to sum  the models are reported in
the  bottom row of the table. 

In Table 2 we  show  the modelling  of galaxies 0211-122 and 0406-244. 
The line fluxes are referred to \Ha=1 for 0211-122 and to \Hb=1 for 0406-244,
following the Humphrey et al. notation. 
 The \Hb flux is given as an upper limit for 0211-122. Considering that the calculated  \Ha/\Hb line ratio is $\sim$ 3 and the lines are referred to \Ha=1,
 \Hb results $\sim$0.33, which is within the upper limit (0.5). Moreover, the spectrum is rich
in lines from different elements in different ionization levels which can constrain the models.
We focus on the [OII]3727+/[OIII]5007+ and CIV1549+/CIII]1909+/CII]2326+ line ratios  (The +
indicates that the doublet is summed)
We could not find any model showing  \Ly/\Ha as low as 0.2.
By a shock dominated model with \Vs=2000 \kms and \n0=1300 \cm3 it is possible to obtain \Ly/\Hb=7 (as for the 0406-244 galaxy)
 and even less. But in this case
HeII 1640/\Hb =9 and all the other line ratios =0  destroy the good fit obtained by the radiation dominated model.
High FWHM
 are adapted to the AGN broad line-emission region (BLR) ($\geq$ 1000 \kms) rather than  to the
narrow line-emission  region (NLR). Forbidden lines   such as [OII] 3727 are generally collisionally deexcited
in the BLR due to the high density. 
Low \Ly/\Ha  perhaps indicates strong absorption by dust in the nebula.
The results  presented in Table 2  show  that the lines come from
different  clouds  within the galaxy. The highest ionization level lines are stronger
where collisional ionization dominates. HeII 1640, which is a recombination line,
increases with  both  the temperature of the gas and  the photoionization flux.

For 0211-122
we  group the shock velocities into two prototypes. \Vs $\sim$ 1000 \kms representing high velocity shocks
fits the UV lines, while
 \Vs $\leq$ 500 \kms   reproduces   the lines in the optical range.
The  results of selected models    appear in cols. 4, 5  and 6.
The weighted sum  of the line fluxes appears in col. 7  of Table \ref{tab2},
Models MH1 and MH2 which refer to \Vs = 1000 \kms and 500 \kms, respectively, are
shock dominated (SD, F=0), while  model MH3 with \Vs=300 \kms is radiation dominated (RD) +shock.
Model MH2 is included because, overpredicting the CIV and [OII] 3727 lines,   improves the approximation  of the
averaged  spectrum. 
 O/H is solar and  N/H  seems lower than solar by a factor of $\sim$ 3. We refer to the solar abundances
by Allen (1976). They are discussed by Contini (2014a).
The average model overpredicts the OIV]/\Ha, NIV]/\Ha and CIII]/\Ha multiplet ratios to \Ha
because the calculations account for the sum of all the multiplet terms. We do not know which terms
are included in the observation data.
 Different geometrical thickness of the clouds $D$  indicate fragmentation  by  shock in a  turbulent regime.
 The 0406-244 spectrum is satisfactorily reproduced by the weighted sum (av in column 11) of models MH4,
 dominated by the AGN radiation, and MH5, dominated by the shock. The relative weights were calculated
phenomenologically in order to obtain the best approximation to the  spectrum, even if some line ratios
(e.g. [NeV]) are less fitted.
Also for this galaxy  a SD  model (MH5) corresponds to the highest \Vs (700 \kms). 
More spectra of   selected galaxies from the Humphrey et al sample are shown and modelled in Table 3.

The FWHM of the line profiles range between 600 and 900 \kms. They are not reported in the table for sake  of space.
In Table 3 we refer all the spectra to \Hb=1
in order to  easily compare observed line ratios to model results from different galaxies. 
 The spectrum presented for 0529-549 in Table 3 shows a few line ratios. Most are upper limits.
Both  [OIII] 5007+/\Hb line ratios calculated  by  high \Vs and  low \Vs models reproduce satisfactorily the data.
[NII]/\Hb and [SII]/\Hb depend strongly on the N/H and S/H relative abundances, respectively.
So, an averaged model is not significant.
The modelling of  0828   suggests that  carbon should be depleted by a factor of $\sim$ 5 from the gaseous phase, 
most probably  trapped into grains 
 throughout the nebula represented by model MH8.
Neon cannot be  included into  dust grains because of its atomic structure, then
we  determine the relative weights of the averaged model av  by fitting  the [NeV]/\Hb line ratio.
Model av (col. 8) shows that, in contrast to our previous hypothesis, C/H solar is suitable to
the high velocity gas component of  0838.
The observed lines of  1138-262 are few and the oxygen  lines are well approximated  by both the AGN and SB dominated models,
therefore  the relative weights in  an eventual average spectrum are less constrained.
Galaxy 4C+23.56 spectrum shows a non negligible [NeV], while 
model MH13 yields [NeV]=0, therefore the SB dominated model MH12 is more appropriated. 

 The main conclusions by Humphrey et al. are in agreement with our modelling. Namely, 
1) AGN photoionization with U varying between the objects  is generally indicated. 
However, we find an SB contribution  to 1138-262 and 4C+23.56 spectra;
2) single-slab photoionization models are unable to  reproduce satisfactorily the high and low ionization 
lines simultaneusly;
3) shocks alone do not provide a valid explanation but must be added to a photoionization source;
4) N/H is nearly solar for all objects 
and does not vary as much as  by a factor of 2-3;
5)  The  line FWHM and  the ionization state are  the result of interaction between the clouds
  in terms of shocks, when the radiation from the AGN or the SB is absent.

\subsection{HDFX 28, a spiral galaxy at z=2.011 from the observations of Dawson et al. (2003)}

 In search for type 2 QSRs, HDFX 28 was
 identified as a face-on spiral galaxy at z=2.011. 
Initially, it was considered  as an extended  radio source  powered by star formation (Richards 2000), 
but the hard X-ray source and the spectroscopy demonstrated that this galaxy
contains an obscured type 2 AGN (Dawson et al and references therein).
The galaxy was observed by ISOCAM and Chandra.
Table \ref{tab4}  shows that  a SB dominated model  (MDS) roughly  reproduces the data.
Dawson et al spectrum  contains too few narrow lines, [NII] 6548, 6548 and \Ha  to
 provide a consistent modelling of the low \Vs gas.
On the basis of  [NII] 6548+6584/\Ha, we can predict that, 
 if the reddening is negligible and \Ha/\Hb=3, [NII]/\Hb= 4.11, which indicates
for a SB model  U$\leq$0.1, \Ts=1-4 10$^4$K, \Vs=220 \kms, \n0=200 \cm3, $D$=10$^{19}$ cm and
solar abundances. For an AGN model, such line ratios 
can be obtained with  N/H solar, \Vs=100-250 \kms and $F$$\leq$ 10$^9$ photons cm$^{-2}$ s$^{-1}$ eV$^{-1}$ at the Lyman limit ,
and for \Vs=500 \kms and  $F$ $\leq$ 10$^{10}$ photons cm$^{-2}$ s$^{-1}$ eV$^{-1}$ at the Lyman limit.
Such [NII]/\Hb can be also obtained by a SD ($F$=0) model   with \Vs=300 \kms 
The preshock densities are 100-300 \cm3 and the geometrical thickness of the emitting clouds
are between 10$^{17}$ and 10$^{19}$ cm.
In conclusion, the [NII]/\Hb line ratio alone cannot constrain the narrow line model.

 In Fig. 2 we present the modelling of the continuum SED. The  data are given by Dawson et al
2003, table 1).  The continuum SED calculated by the code SUMA shows two curves for each model.
One represents the bremsstrahlung from the gas and the second one, in the IR, represents reprocessed
radiation from dust.
The data in the  radio - UV - soft-X-ray range are well fitted by the high \Vs, SB dominated model MDS.
Harder X-ray data in Fig. 2  could   be reproduced by a model calculated by  a shock velocity even higher than 1500 \kms,
as suggested by the FWHM of some line profiles.
The  AGN and the SB dominated models calculated by \Vs=220 \kms, which were  predicted on the basis of 
the [NII]/\Hb line ratio, 
  reproduce the data in the optical range, in agreement with the \Ha narrow line FWHM. 
The modelling of the  HDF galaxy continuum SED constrains the choice of the low velocity model,
but it cannot select   between an AGN or a SB.

\begin{figure}
\centering
\includegraphics[width=9.0cm]{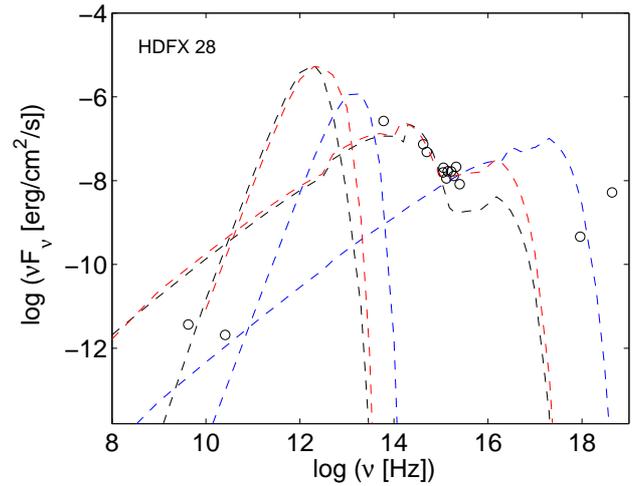}
\caption{Modelling the   continuum SED  observations   Dawson et al. (2002, table 1 and references therein).
Blue lines:the shock dominated model MDS model;  black lines : the AGN radiation dominated  model calculated
by \Vs=220 \kms, \n0=200 \cm3,  log$F$=9, and solar abundances; red lines : the SB radiation dominated
model calculated by \Ts=3 10$^4$K and $U$=0.1.
}
\end{figure}

\subsection{Spectrum from CXO 52 by Stern et al. (2002) observations}

 CXO 52 was identified   as an optical, colour selected, high redshift source (Stern et al 2002c) 
and  by optical follow up  
of X-ray sources in the Lynx field (e.g. Stanford et al 1997) .
Stern et al (2002) report on observations of a type 2 quasar at z=3.288 that was identified
as a hard X-ray source by Chandra X-ray observatory.
The optical spectrum of CXO 52 was obtained by Keck I telescope.
 Line profile FWHM  are  broad ($\geq$ 1000 \kms). A shock dominated model
with \Vs=1500 \kms and   pre-shock density  \n0=110 \cm3,
  resembling those found in other galaxies at  similar redshifts, fits most of the UV lines
(MSS Table 4).  The \Ly/\Hb is 
overpredicted by a factor of 1.6.
OVI1035/\Hb is underpredicted and NV 1240/\Hb is well reproduced.
 The maximum temperature of the gas downstream, near the shock-front
depends on the shock velocity (Sect. 2)  therefore, 
MSS model fits 
 NV/\Hb   because NV is a  high ionization level line. 
OVI  which shows a FWHM =2640 \kms cannot 
be well reproduced by model MSS. 

The \Hb and the forbidden
lines [OIII]5007+  correspond to lower FWHM (170-430 \kms). 
 The lines are emitted from  clouds within the galaxy with different \Vs. 
A  radiation dominated model (MSR) with \Vs=200 \kms and \n0=350 \cm3  reproduces the UV lines (except OVI and NV) 
within a factor of 2 and explains  the high  [OIII]5007+4959/\Hb.
The flux from the AGN corresponds to a  Seyfert 2.
The results of the two models  are not summed up because
they   show very different gas conditions in (most probably) well separated 
 regions throughout the galaxy.

 To cross check the models we  compare the bremsstrahlung calculated  by models 
MSS and MSR with the data presented by Stern et al (in their table 1) in Fig. 3.
The datum at 2-10 keV is well reproduced by model MSS, calculated by a high shock velocity.
The bremsstrahlung maximum depends on the maximum temperature of the gas
downstream.  Dust grains  can be completely destroyed by sputtering  at high
shock velocities ($>$  1000 \kms). 
The grains in the nebulae represented by models MDS and MSS are characterised by  a relatively  large initial radius  
(\agr=1 \mum) in order to survive
to strong sputtering  throughout a high velocity  shock-front. However, they
  can be completely destroyed  by sputtering and evaporation, as happens for model MSS,  
depending on the cooling rate downstream.
The  relatively low \Vs which characterises  the gas conditions  in model MSR yields  
grain survival. 
The bremsstrahlung calculated by model MSR, which corresponds to
gas  ionised and heated  by the flux from the AGN (+shocks),  roughly fits the data in the UV-optical and radio ranges.
The reprocessed radiation by  dust in the IR is constrained    by only one datum  which  leads to
 $d/g$  = 10$^{-4}$ by mass.

\begin{figure}
\centering
\includegraphics[width=9.0cm]{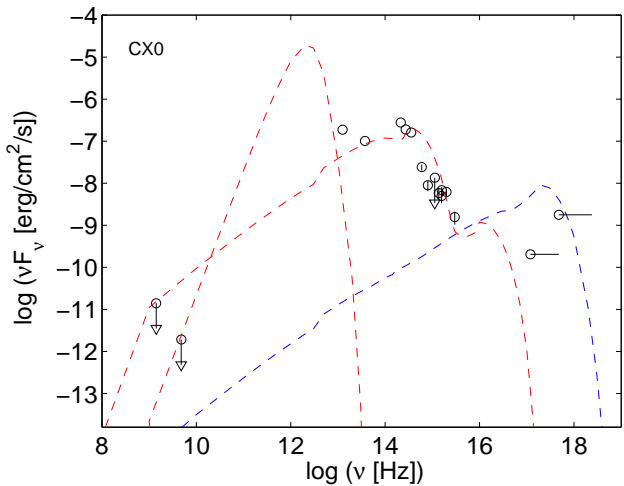}
\caption{Modelling the   continuum SED  observations  Stern et al. (2002, table 1 and references therein).
Blue lines:the shock dominated model MSS model; red lines : the radiation dominated  model MSR;
}
\end{figure}

\subsection{The spectrum observed from Q2343-BX418 by Erb et al (2010)}

In  column  9 of Table 4 we present the observed  line ratios for
 BX418 galaxy at z=2.3, defined by Erb et al., as "one of the youngest and lowest stellar mass
continuum-selected z$\sim$2 galaxies".

The spectral lines show different FWHM for different wavelengths decreasing from
$\sim$ 840 \kms for \Ly to 640 \kms for HeII 1640, down to $\leq$ 260 \kms for lines
in the optical-near-UV range (Table 4, col 10).
The  observations show that different components contribute
to different lines.
Although   permitted line fluxes  (e.g.  \Ly and HeII)  depend strongly on the
photoionizing radiation  reaching the nebula,  they  also increase   with the temperature
of the emitting gas (see Sect. 3.8). Therefore, models accounting for   shock+ photoionization are used.
The models are presented in Table 4. We first investigate whether an AGN (hidden?) is present.
The high \Ly and HeII 1640 FWHM suggest shock velocities of $\sim$ 850 and $\sim$ 600\kms,
respectively. They are accompanied by a AGN flux characteristic of Seyfert 2 galaxies.
The densities are high throughout the nebula  to reduce the primary and secondary
fluxes rapidly enough. The temperature drop will lead to very low
line intensities in the optical range. Actually, broad optical lines were not observed.
We could not obtain acceptable  results with models corresponding to outflow of the gas,
on the contrary,   gas inflow  towards the AGN is indicated (models ME1-0 and ME2-0).
We  reproduced the near UV and optical line ratios by models corresponding to 
the NLR conditions. Models ME3 and MESB roughly fit the [OII]/\Hb and [OIII]/Hb line ratios.
ME3 refers to an AGN, while MESB  refers to a SB radiation source.
Both shows outlow of gas from the galaxy and are similar to those found in other galaxies
at those redshifts. We suggest that all  the  gas conditions represented by the models in Table 4
are present within BX418,
contributing to the single lines in the  spectrum.  The models are not summed up,
in order to show  each contribution to each  line.
 BX418 has been placed among metal-poor galaxies (Erb et al). However, the low metallicity (0.05 solar)
was calculated by Erb et al by the strong-line metallicity diagnostic as well as by the direct methods.
They found 12+log(O/H)=7.9$\pm$0.2, while
the results of detailed modelling yield   solar 12+log(O/H)=8.812.
The discrepancy between  metallicities  calculated by direct methods and modelling has been discussed
by Contini (2014b).

\subsection{Spectra by Vernet et al (2000) at z$\sim$ 2.5 and the diagrams}

Vernet et al (2001)  present  spectropolarimetry of powerful extragalactic radio sources
at z$\sim$2.5 by Keck II telescope, "when quasars were much more common than now".
Although dust-reflected quasar light dominates the UV continuum, they did not exclude
  star formation.  
Vernet et al.    comparing the data throughout the NV/HeII vs NV/CIV diagrams by Hamann \& Ferland (1993) 
 could not easily explain   the  observed NV/CIV and NV/HeII line ratios.
In  Sect. 4 we will discuss Vernet et al and other survey  observations throughout  the HeII/CIV vs \Ly/CIV diagrams,
where  the  data are compared with model results  calculated  by SUMA.
 UV line ratios are generally referred to CIV  which is
one of the strongest lines. Moreover, we have found that the N/H relative abundances throughout galaxies are scattered
by a factor of $\sim$ 10 (Contini 2014a, fig. 5), therefore we suggest that  NV lines  are less 
adapted to diagnostic diagrams.

 We  present in Table \ref{tab5} the results of  shock+photoionization models for  some of the Vernet et al spectra. 
  4C+40.36 and 4C+48.48  were excluded because the spectra are most probably  blended by a foreground HII galaxy at z=0.404
and at z=0.684, respectively.
 The modelling results indicate that composite models referring to the SB+shocks, to the AGN+shocks
should be adopted  and that different physical conditions from different regions contribute to the spectra.
 All the  models used to reproduce  Vernet et al  data in Table 5  show relatively low \Vs and \n0,
which are reflected in  low SFR.
Vernet et al found a SFR range between $\sim$ 60 \Moy to $\leq$ 2 \Moy.
They claim that in 0731+438 the only object where reprocessed radiation from AGN does not dominate the continuum
(for the other objects starlight conributes less than half of the UV continuum,  see Vernet et al 2001, table 5) 
and assuming E(B-V)=0.1 and a stellar population age of 10$^9$ yr, Vernet et al calculate SFR
between 30 and 120 \Moy.
Our calculations give SFR $\sim$ 128 \Moy for 0731+438, a minimum SFR $\sim$ 24.4 for 4C+23.56a, etc. (see Table 5)

\begin{figure}
\centering
\includegraphics[width=9.0cm]{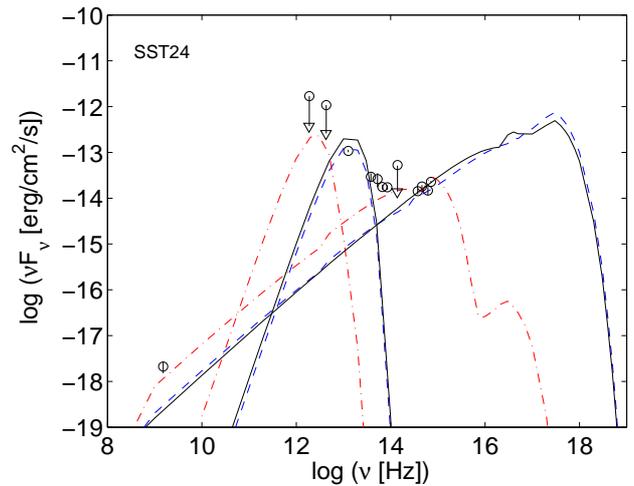}
\caption{The  continuum SED modelling of  Dey et al. data.
Blue lines:the AGN model; red lines : the SB model; black lines: the shock dominated model
and reradiation by dust at 1000 K}
\end{figure}

\begin{figure}
\centering
\includegraphics[width=9.0cm]{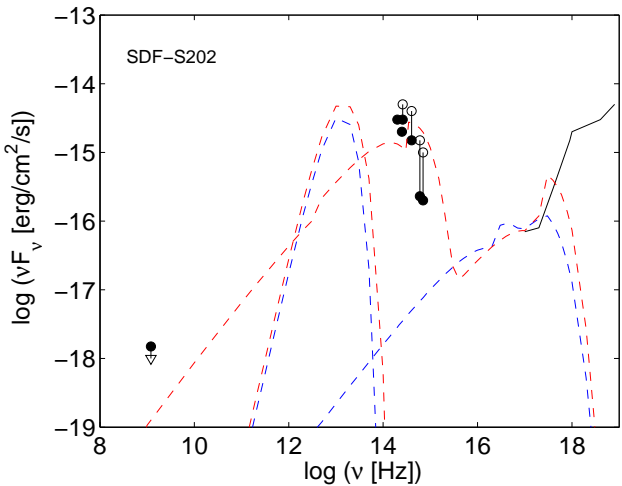}
\caption{The  continuum SED modelling of SFD-S202.
The data are  adapted from Norman et al. (2002, fig. 6)
Blue lines: the shock dominated  model MNS; red lines : the AGN  model MNR.
black solid line : the X-ray data}
\end{figure}

\subsection{Line and continuum observations of an extended nebula by  Dey et al. (2005)}

Massive ellipticals  at high z ($\geq$3) are surrounded by large (100-200 kpc) luminous
($>$10$^{43}$ erg s$^{-1}$ ) \Ly nebulae and small embedded star forming objects.
Investigating  ($>$100 kpc) halos and environments of luminous radio galaxies affected by AGNs 
and ejecta  of similarly large \Ly  nebulae, 
 Dey et al (2005) reported on observations  of a very large, spatially extended (160 kpc)  \Ly emitting nebula at z=2.656.
Spectroscopic observations of SST24 J1434110+331733  were obtained by the Low Resolution Imaging Spectrometer
(LRIS) on the Keck I telescope. 

 The spectroscopic observations reveal  very complex regions within the extended nebula, 
including several continuum sources.
Dey et al suggest that many active galaxies are forming in that region
which lead to three main contributions to the continuum spectrum. 
Galaxy A shows interstellar absorption lines and no \Ly emission; 
 the spectrum of the second 
 continuum source is quite red and  shows weak
CIV and CIII] emission at z=2.656. Star formation is suggested because the source is close to MIPS 24 \mum location.
The third continuum source lies in the centre of the \Ly nebula associated to CIV and HeII.
 The velocity structure of the nebula deduced from the line profiles shows that the velocity varies uniformly
across the central region. If due to gas kinematics, the variation can be due to infall,
outflow, or rotation.

 We refer to  the line and continuum observations  by Dey et al. 
 The results of  modelling are presented in Table 6.
On the basis of the  observed lines corresponding to broad profiles, we have run
models with \Vs$>$ 1000 \kms. 
We find a large contributions to the line fluxes  from high velocity  (\Vs=1700 \kms) shock dominated gas (models MDOS)
 and photoionized  by an AGN (model MDR). Moreover,  the contribution  from gas with \Vs=300 \kms
and photoionised by a SB  is evident (MDSB).
The high velocities are at the limit of those  observed  in AGN NLR. The AGN flux is characteristic of Seyfert 2.
The starburst temperature which leads to a  satisfactory fit of the line ratios (2.5 10$^5$ K) is as high as 
that found in galaxies showing  activity (Contini 2014b). Also
 the ionization parameter is high.
The  gas throughout the SB shows  "normal physical conditions"   in the emitting nebulae  of galaxies at such redshifts.
However, to have a good fit of the NV 1240 line, we have adopted N/H lower than solar by a factor
of $\sim$ 10 in the shock dominated gas. This indicates that  it  is located in a different  region of the galaxy.
The   weighted sum of the  calculated spectra  (av) appears in  Table 6 col. 6. 
We suggest that the  $\lambda$ 5084 line refers to  scandium ($^4$F$_{3/2}$-$^4$F$^0_{9/2}$ , White 1934, fig.14.7). 
Scandium can be present is SN ejecta (e.g. Bose et al. 2015).

 Fig. 4 shows that the continuum SED   calculated by the  models selected by fitting the line ratios
  reproduce the observed continuum data
presented by Dey et al. (2005, fig.2).  
Bremsstrahlung from  gas heated by the strong shock shows a maximum at high frequency (Contini, Viegas, Prieto  2004).
X-ray emission is predicted.
 The mass of the emitting gas (6 10$^{12}$ \msol)
was calculated by Dey et al. 
The mass of dust can be calculated by  $d/g$  = 0.003 by mass which results from the 
fit of the continuum SED in the IR.

\begin{figure}
\centering
\includegraphics[width=9.0cm]{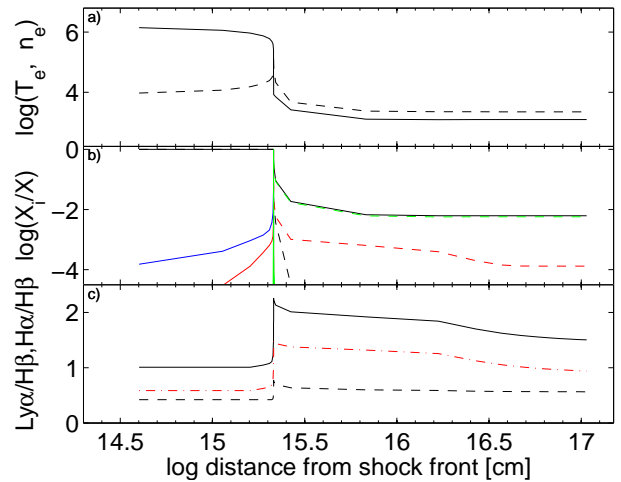}
\caption{The profiles of the physical quantities throughout the emitting nebula for
the SD  model MF0. The shock-front is on the left.
Top : electron temperature (solid black), electron density (dashed black).
Middle : H$^+$/H (solid black), He$^{++}$/He (dashed black), C$^{2+}$/C (dashed red),
C$^{3+}$/C (solid red), N$^{4+}$/N (solid blue), O$^+$/O (dashed green), O$^{2+}$/O (solid green).
Bottom : log (\Ly/\Hb) (solid black), log(\Ha/\Hb) (dashed black) and log(\Ly/\Ha) (dot-dashed red).
}
\end{figure}

\begin{figure*}
\centering
\includegraphics[width=9.0cm]{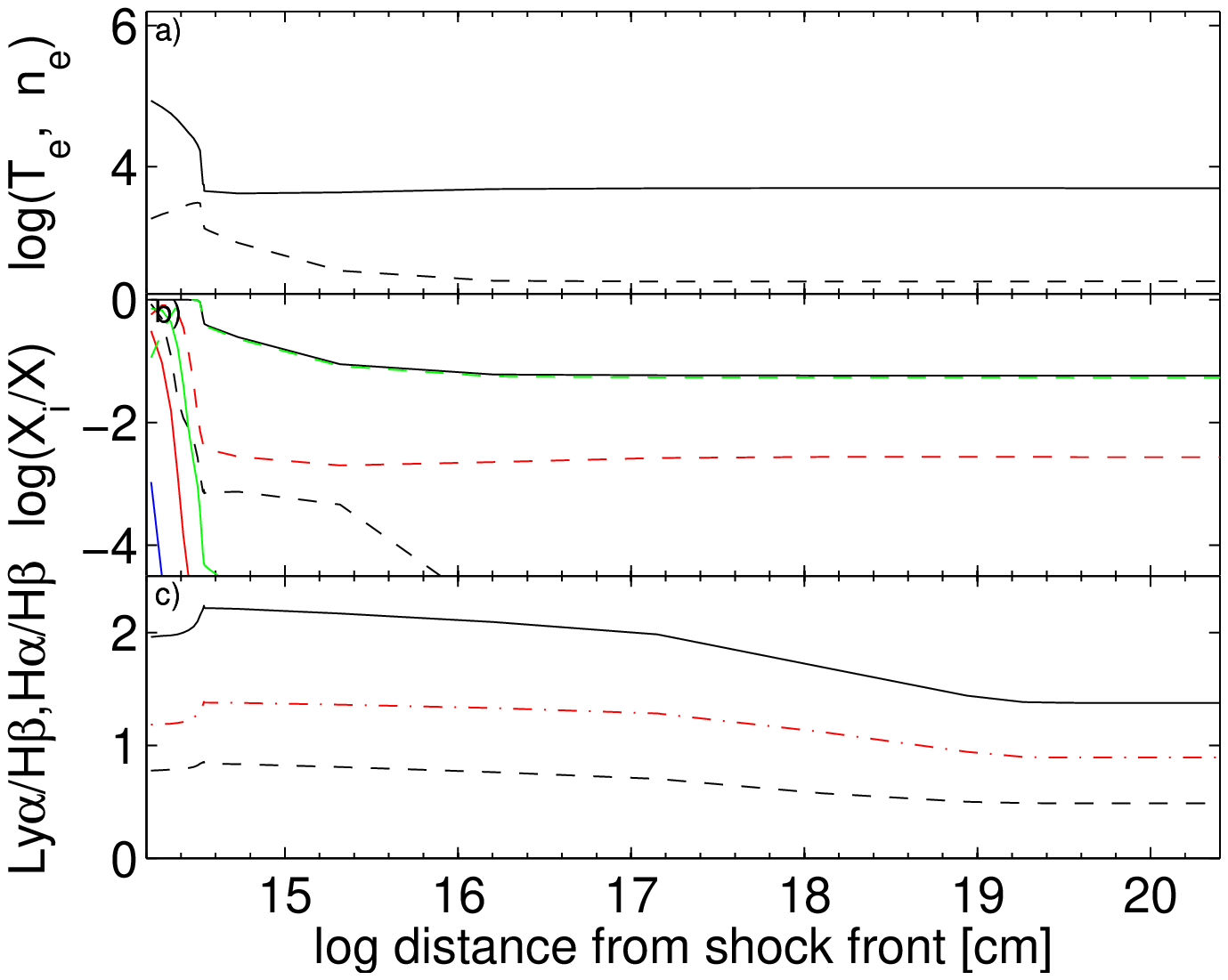}
\includegraphics[width=9.0cm]{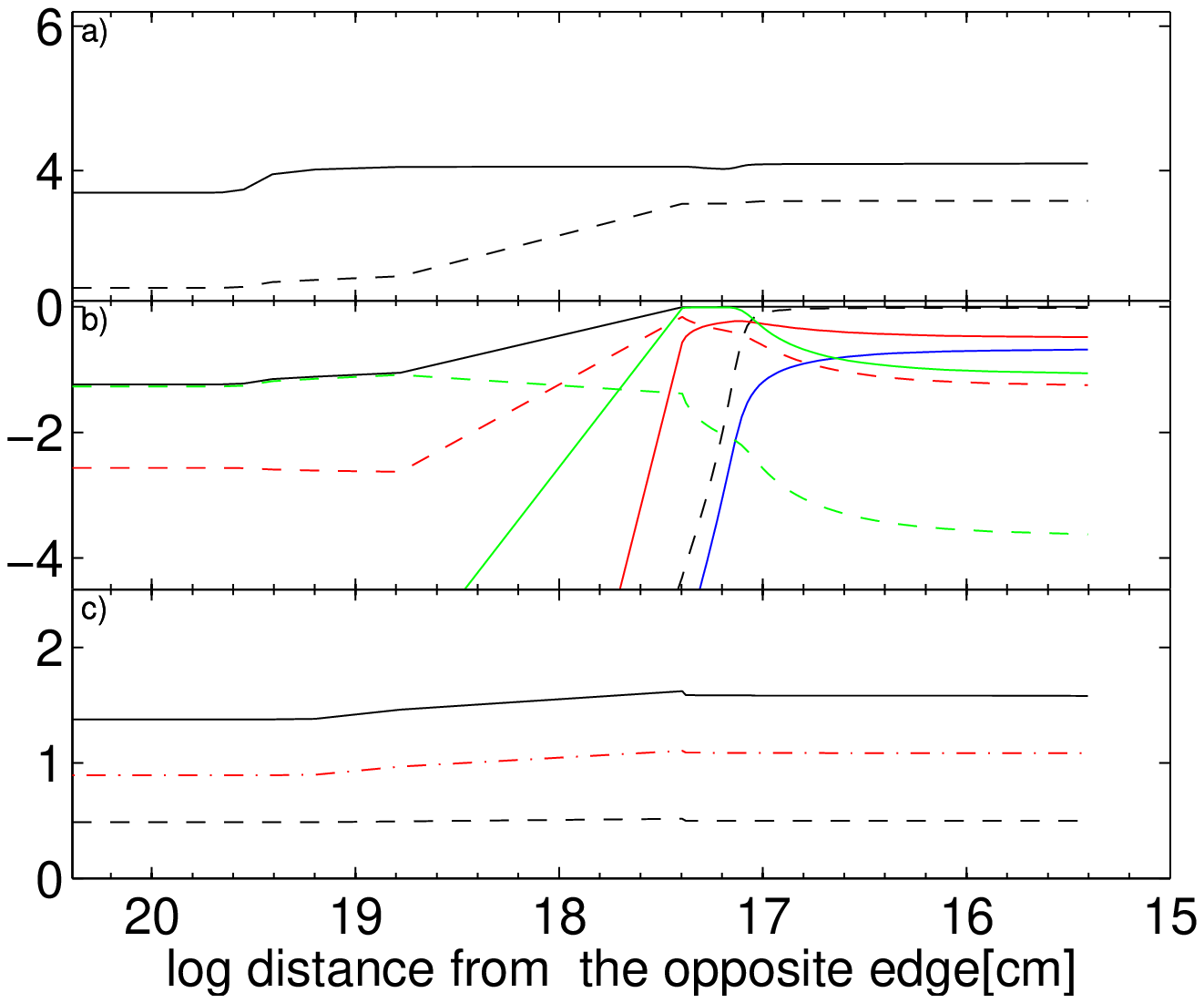}
\caption{Same as in Fig. 6 for  model MFSB}
\end{figure*}

\subsection{A classic type 2 QSO at z=3.7 : CDF-S202 observations by Norman et al (2002)}

Norman et al observed the  AGN  detected at z=3.7  by the Chandra Deep Field-South 1 Ms exposure.
It is  a distant type 2 AGN with UV   showing FWHM of $\sim$ 1500 \kms.
The optical spectrum was obtained with the multislit mode on the FORS1 on the ESO/VLT-ANTU.
Detailed modelling  (Table 6, cols.  9, 10) indicates that a shock dominated model (MNS) best 
fits the lines within a maximum  discrepancy  factor of 
2.3 for HeII 1640 (Table 6 last column).
This model calculated by  \Vs=1600 \kms would provide  OVI 1035/\Hb $\sim$ 200.
Norman et al  found a high metallicity, while we suggest  N/H by a factor of 1.5  lower than solar
 and C/H  about solar, considering the uncertainty of the line fluxes.
Referring the line to CIV=10 instead of  \Hb,  as it is generally done for UV
spectra, we would obtain an acceptable fit of all the lines.
The radiation (AGN+shock) dominated model (MNR) calculated by  the high \Vs  reproduces
the line ratios roughly.
 In Fig.  5  we present the modelling of the continuum SED. The data were adapted from Norman et al. fig. 6.
The continuum calculated by the  radiation (+shock) dominated model MNR reproduces the data in the optical- UV 
better than the shock dominated model MNS, so the MNR model should be accounted for, but  with a low relative weight.

\begin{table*}
\begin{center}
\caption{Modelling  the  corrected line ratios to \Hb=1 observed  by  Finkelstein et al (2011)}
\tiny{
\begin{tabular}{lcccccccccccc} \hline  \hline
\  ID  &  z     &SFR   &    \Ly& \%  &   NV  &   CIV   & HeII   & [OII]  &  [OIII] & \Ha    &   [NII]   \\
\      &        &\Moy  &  1215 &  -   & 1240  & 1550    & 1640   & 3727+  &  5007+  & 6563   & 6584     \\ \hline
\ EGS7 & 0.2440 &0.4   & 17.53 &15.9  &  -    &   -     &  -     & 2.14   &  5.65   & 3.     & 0.134    \\
\ MSB7 &        &      & 24.3  &-    & 0.0014& 0.34    & 0.078  & 2.3    &  5.6    & 2.93   & 0.16     \\
\ MAG7 &        &      & 28.6  &- & 0.0009& 0.15    & 1.22   & 2.     &  6.4    & 3.3    & 0.18     \\    
\ EGS8$^*$ &0.2395 &0.035&  16.7&32.8  & -     &  -      &  -     & 2.47   &  3.7    & 3.     & $<$0.03  \\
\ MSB8 &         &     &23.8  &-  & 0.0024& 0.28    & 0.0037 & 2.     &  3.5    & 2.95   & 0.1      \\
\ MAG8 &         &     &28.6  &-  & 0.0005&0.14     &0.89    & 2.     &  3.46   & 3.34   & 0.1      \\ 
\ EGS10 & 0.2466 & 0.12&33.9  &16.7  &   -   &   -     &   -    & 3.06   &  4.37   & 3.     & $<$0.06  \\
\ MSB10 &        &     &24.6  &-  &  0.016&  0.46   & 0.044  & 3.2    &  4.5    & 2.94   & 0.12     \\          
\ MAG10 &        &     &28.   &-  & 0.001 & 0.029   & 1.1    & 3.1    &  4.65   & 3.2    & 0.14     \\
\ EGS11 & 0.2524 &0.059& 41.1  &35.3  &   -   &   -     &   -    & 2.78   &  1.53   & 3.     & 0.84      \\
\ MSB11 &        &     &24.5  &-  & 0.1   & 0.3     & 0.0026 & 2.3    &  1.5    & 2.96   & 0.87     \\
\ MAG11 &        &     &26.5  &-  & 0.0027& 0.1     & 0.67   & 2.5    & 1.57    & 3.3    & 0.9      \\  
\ EGS12 & 0.2515 &0.062& 48.7  &20.8  & -     &  -      &   -    & 2.8    & 2.7     & 3.     & 0.38    \\
\ MSB12 &        &     &24.5  &-  & 0.067 & 0.42    &  0.01  & 2.9    & 2.63    & 3.     & 0.46   \\
\ MAG12 &        &     &28.   &-  & 0.0027& 0.28    &  0.8   & 2.5    & 2.8     & 3.3    & 0.37   \\
\ EGS13 & 0.2607 &0.46 &  17.96 &21.  &   -   &   -     &   -    & 2.8    & 2.26    & 3.     & 0.43   \\
\ MSB13 &        &     &24.5  &-  & 0.067 & 0.42    & 0.0092 & 2.9    & 2.45    & 2.95   & 0.47   \\
\ MAG13 &        &     &31.54 &-  & 0.017 & 0.25    & 0.73   & 3.     & 2.35    & 3.3    & 0.46   \\   
\ EGS19 & 0.2666 &0.11 & 33.96 &21.9  & -     &  -      &  -     & 2.4    & 1.23    & 3.     & 0.86   \\
\ MSB19 &        &     & 24.4  &-  &0.1    & 0.31    & 0.0046 & 2.36   & 1.3     & 3.     & 0.86   \\
\ MAG19 &        &     &26.4  &-  & 0.0027&0.1      & 0.61   & 2.5    & 1.25    & 3.     & 0.96  \\ 
\ EGS20$^*$ & 0.2680&0.073&11.57&27.2 &  -    &  -      &   -    & 3.04   & 4.2     & 2.9    & $<$0.137 \\
\ MSB20 &        &     &24.3   &- & 0.024 & 0.46    & 0.015  & 2.96   & 4.12    & 2.93   & 0.14 \\
\ MAG20 &        &     &28.    &- & 0.001 & 0.29    & 1.1    & 3.     & 4.5     & 3.     & 0.14 \\ 
\ EGS21 & 0.2814 &0.06 &   46.4   &21.6 &   -   &  -      &   -    & 1.98   & 0.415   & 3.     & 0.12-1.4\\
\ MSB21 &        &     &32.    &- & 0.063 & 0.155   & 0.0015 & 2.2-1.8& 0.47-0.4& 3.3    & 0.13-1.2\\
\ MAG21a&        &     &30.5   &- & 0.002 & 0.3     & 0.27   & 1.8    & 0.45    & 3.46   & 1.5 \\
\ MAG21b&        &     &28.3   &- & 0.0027& 0.14    & 0.31   & 2.1    & 0.42    & 3.4    & 0.15 \\  
\ EGS22 & 0.2828 &0.12 &  242.   &35.9 & -     &  -      &  -     & 3.17   & 0.74    & 3.     & $<$0.14 \\
\ MSB22 &        &     &25.5   &- & 0.023 & 0.9     & 0.011  & 3.     & 0.75    & 2.96   & 0.13 \\
\ MAG22 &        &     &27.9   &- & 0.002 & 0.22    & 0.5    & 3.2    & 0.72    & 3.3    & 0.12 \\
\ EGS23 & 0.2865 &0.42 & 18.4   &15.8 &  -    &  -      &  -     & 2.65   & 3.7     & 3.     & 0.32 \\
\ MSB23 &        &     &24.2   &- & 0.045 & 0.55    & 0.012  & 2.65   & 3.74    & 2.93   & 0.31 \\
\ MAG23 &        &     &27.1   &- & 0.004 & 0.3     & 1.0    & 2.60   & 3.88    & 3.2    & 0.37 \\
\ EGS25 & 0.3243 &1.06 & 6.5    &16.6 & -     &  -      &   -    &  1.76  & 5.94    & 3.     & 0.16 \\
\ MSB25 &        &     &24.    &- & 0.04  &  0.68   & 0.023  &  1.8   & 6.2     & 2.93   & 0.15 \\
\ MAG25 &        &     &28.9   &- & 0.0005& 0.11    & 1.13   & 1.8    & 5.95    & 3.3    & 0.14 \\ \hline   
\end{tabular}}

$^*$ not Balmer corrected because the observed \Ha/\Hb $\sim$ 3.

\end{center}

\end{table*}

\subsection{Spectrum of the Lynx arc at z=3.357  by Fosbury et al (2003)}

The Lynx arc (z=3.357)is characterised by a very red R-K colour  and strong, narrow emission lines.
Fosbury et al claim that analysis by HST WFPC2 imaging and Keck optical and IR spectroscopy indicate 
 a HII galaxy
magnified by a factor of 10 by a cluster environment
including  stars with \Ts$\sim$ 10$^6$K,
 $U$ $\sim$ 1  and low metallicity (0.05 solar).
In  Table 6, col 11  the observed spectrum of the Lynx arc 
presented by Fosbury et al (2003, table 2) is reported.
The spectrum shows a relatively large number of line ratios including
[OII]/\Hb and [OIII]/\Hb in the optical range, reproduced
by Fosbury et al  by a pure   photoionization model
with \Ts=8 10$^4$ K  log$U$=-1 and Z=0.05Z$_{\odot}$,
 which was suggested by comparison with  synthetic SEDs calculated by STARBURST99 (Leitherer et al 1999).
The  FWHM of the  profiles are $\leq$ 100 \kms for 
 lines in the optical  and higher for those in the UV.
We  searched by modelling the spectrum for a (hidden) AGN, but 
 AGN dominated models lead to HeII/\Hb $\geq$1, higher by a factor $\geq$ 10 than observed.
The shock dominated model MF0 (Table 6, col. 12) calculated by a relatively high shock velocity (320 \kms)
and high preshock density (\n0=2000 \cm3) fits most of the UV line ratios   but underpredicts 
 [OIII] 5007+/\Hb  by a factor $\geq$ 10.
 A  SB radiation dominated model MFSB, calculated by   \Ts =1.1 10$^{5}$K, $U$=0.3, 
$D$=4.9 10$^{20}$ cm, \Vs=100 \kms,  \n0=300 \cm3 and O/H $\sim$ 0.5 solar,
reproduces satisfactorily most of the  line ratios from relatively low  ionization levels. 
Model MFSB which accounts for photoionization by a SB  + shocks represents gas ejected outwards, 
while for model MF0 the direction of the emitting gas is ambiguous.

Finally, we show in Figs. 6 and 7 the profiles of the physical  parameters throughout
the emitting nebula for models MF0 and MFSB, respectively, in order to understand the results.
In Figs. 6 and 7  the electron temperature and electron density  
appear  in the top diagrams. 
In Fig. 7 the nebula is devided into two halves in order to obtain a comparable view 
 at the two edges. Logarithmic symmetric scales are  used for the X-axis.
The shock front is on the left of the left panel  and the right edge of the right panel
 is reached by the radiation flux from the  SB.
In the middle diagrams of Figs. 6 and 7 the  fractional abundances of the ions corresponding
to significant lines are shown.
 The bottom diagrams are dedicated to the \Ly/\Hb, \Ha/\Hb and \Ly/\Ha line ratios (in logarithm).
Figs 6 and 7 show that \Ly/\Ha follows  the  H$^+$ recombination trend.
The two sides of the nebula (shock dominated at the left and radiation dominated at the right of Fig. 7)
are bridged by diffuse secondary radiation. In fact, the temperature 
is $\sim$ 10$^4$K even in the extended internal region of the nebula.
Most of the C, N, O lines  come from the radiation dominated side of the nebula.

\subsection{Finkelstein et al survey}

 Finkelstein et al (2011) present in their table 1 the measured line fluxes for a sample of galaxies
at 0.24 $\leq$ z $\leq$ 0.324 by spectroscopy of LAE using the Hectospec spectrograph with the
6 m MMT telescope.

We report in Table 7 the results of modelling the observed line ratios and we add the
 UV lines NV, CIV and He II consistently calculated by the same models. These lines
which are not  observed by Finkelstein et al   will be  discussed  in the following.
In Table  7 the observed line ratios (EGS7-EGS25, corrected by  \Ha/\Hb =3) are followed
in the next rows by the
results of SB models (MSB7-MSB25)) and AGN models (MAG7-MAG25)).
The input parameters are reported in Tables 8 and 9).

 \Ly lines  are calculated consistently with the
other lines.
 \Ly/\Hb  calculated line ratios (col. 3)  reproduce the data within a factor of 2.
In col. 4 the SFR are indicated for each galaxy. The percentage uncertainty appears in col. 5.
The galaxies in the Finkelstein et al sample  present rather homogeneous
characteristics. The O/H relative abundances are at most solar while the N/H ratios are
lower than solar by a factor $<$ 10. Densities and velocities are relatively low.
The flux from the active nucleus is in the range of that found in low luminosity AGNs.
 The temperature of the stars does not show the maxima corresponding to outbursts (Contini 2014b).
The geometrical thickness of the clouds shows that fragmentation is present in the
SB neighbourhood  rather than in AGNs.

\begin{table}
\caption{Input parameters  for Finkelstein et al. MSB  models}
\tiny{
\begin{tabular}{llccccccccccc} \hline  \hline
\ model &         \Vs  &  \n0  & \Ts   & $U$  &  N/H  & O/H     & $D$   & \Hb \\ 
\      &            1    &  2    &  3      & 4 &  5       & 6       &   7          & 8 \\ \hline
\  MSB7&          100  &  100  &  6.4  & 0.02  & 0.1  &  5.   &  9. &  0.02   \\
\  MSB8&         100  &  100  &  4.2  & 0.03  & 0.2  &  6.6  &  3. &  0.0196\\
\  MSB10&         150  &  100  &  5.9  & 0.02  & 0.1  &  6.0  &  4. &  0.019 \\
\  MSB11&         150  &  100  &  4.   & 0.03  & 0.9  &  6.6  &  4. &  0.027\\
\  MSB12&         180  &  100  &  4.8  & 0.025 & 0.4  &  6.6  &  5. &  0.024  \\
\  MSB13&         180  &  100  &  4.7  & 0.025 & 0.4  &  6.6  &  5. &  0.025 \\
\  MSB19&         210  &  100  &  3.9  & 0.04  & 0.8  &  6.6  &  5. &  0.038 \\
\  MSB20&         220  &  100  &  4.8  & 0.035 & 0.13 &  6.6  &  1. &  0.029 \\
\  MSB21&         170  &  100  &  4.   & 0.015 & 0.1  &  6.6  &  300 &  0.058 \\
\ MSB22&           200  &  150  &  4.2  & 0.015 & 0.06 &  5.2  &  80 &  0.023  \\
\ MSB23&          190  &  100  &  4.4  & 0.034 & 0.2  &  5.6  &  1. &  0.022 \\
\ MSB25&          170  &  100  &  4.9  & 0.034 & 0.15 &  5.   &  1. &  0.023 \\ \hline
\end{tabular}}

1:in \kms ; 2:in \cm3 ; 3:in 10$^4$K ; 4: - ; 5:in 10$^{-4}$ ; 6:in 10$^{-4}$ ; 7:in 10$^{17}$ cm ; 8:in \erg 

\end{table}

\begin{table}
\caption{Input parameters  for Finkelstein et al. MAG  models}
\tiny{
\begin{tabular}{llccccccccccc} \hline  \hline
\ model& z       &  \Vs  &  \n0  & $F$  &  N/H  & O/H     & $D$   & \Hb \\ 
\      &         &   1   &   2   &   3  &  4    &  5      &  6    &  7  \\ \hline 
\  MAG7&0.2440 &   100 &   150 &   7. &   0.2 &    5.6  &   3  &  0.066  \\
\  MAG8&0.2395 &   100 &   150 &   3. &   0.1 &    5.6  &   3 &  0.064 \\ 
\  MAG10&0.2466 &   100 &   150 &   3. &   0.1 &    6.6  &   3 &   0.029\\ 
\  MAG11&0.2524 &   100 &   150 &   1.4&   0.8 &    6.6  &   3 &  0.072  \\
\  MAG12&0.2515 &   100 &   150 &   2.5&   0.3 &    6.0  &   1.4&  0.03   \\
\  MAG13&0.2607 &   120 &   150 &   2.6&   0.3 &    6.0  &   1.9&  0.045  \\
\  MAG19&0.2666 &   100 &   150 &   1.2&   0.8 &    6.6  &   3 &  0.072  \\
\  MAG20&0.2680 &   100 &   150 &   2.9&   0.1 &    6.6  &   1 &  0.03   \\
\  MAG21a&0.2814 &   100 &   150 &  0.77&   0.2 &    6.6  &   1.6&  0.026  \\
\  MAG21b&0.2814 &   110 &   150 &  0.8 &   0.15&    6.6  &   4 &  0.068  \\
\  MAG22&0.2828 &   110 &   150 &   0.86&   0.08&    6.0  &   4 &  0.045  \\
\  MAG23&0.2865 &   100 &   100 &   1.8&   0.3 &    6.6  &   2 &  0.018  \\
\  MAG25&0.3243 &   100 &   150 &   6. &   0.15&    5.8  &   4 &  0.086  \\ \hline
\end{tabular}}

1:in \kms ; 2:in \cm3; 3: in 10$^9$  photons cm$^{-2}$ s $^{-1}$ eV$^{-1}$ at the Lyman limit ;
4:in 10$^{-4}$; 5:in 10$^{-4}$; 6:in 10$^{19}$cm ; 7:in  \erg

\label{tab9}
\end{table}

\subsection{Predicted \Ly and UV lines from galaxies at 0.06$<$z$<$0.9}

\begin{table*}
\caption{The  data observed by Ly et al (2014)}
\tiny{
\begin{tabular}{cccccccccccc} \hline  \hline
\ ID   & zspec & FWHM &  [OII]& \Hb    &[OIII]   & [OIII]     & \Ha  & [NII]& log([OII]/\Hb) &  log([OIII]5007+/\Hb) & log(\RO3)  \\
\       &       & \kms & 3727+ & 4861   & 4363    &5007+       &6563  & 6583 &        -       &               -       &       -               \\ \hline
\ MMT01 & 0.6380& 220.0&  10.58 &  9.21&   2.25   & 73.32     &   0.00& 0.00 &   0.184$_{0.174}^{0.195}$ &0.886$_{0.065}^{0.082}$&1.462$_{0.226}^{0.211}$\\
\ MMT02 & 0.4327& 247.0&  18.20 &  5.76&  0.60    & 25.83     &  0.00 & 0.00 &   0.621$_{0.173}^{0.287}$ &0.644$_{0.044}^{0.040}$&1.549$_{0.394}^{0.282}$\\
\ MMT03 & 0.4809& 256.4&  20.40 &  5.81&  0.95    & 12.23     &  0.00 & 0.00 &   0.735$_{0.200}^{0.211}$ &0.339$_{0.049}^{0.052}$&1.086$_{0.285}^{0.269}$\\
\ MMT04 & 0.3933& 248.6&  49.30 & 21.70&  1.38    &130.05     &  0.00 & 0.00 &   0.355$_{0.009}^{0.026}$ &0.774$_{0.008}^{0.007}$&1.950$_{0.141}^{0.129}$\\
\ MMT05 & 0.3846& 232.1&   3.70 &  7.02&  1.29    & 57.92     &  0.00 & 0.00 &  -0.266$_{0.102}^{0.147}$ &0.921$_{0.034}^{0.029}$&1.659$_{0.192}^{0.158}$\\
\ MMT06 & 0.3995& 260.2&  21.33 &  6.04&  0.83    & 21.97     &  0.00 & 0.00 &   0.734$_{0.061}^{0.069}$ &0.543$_{0.051}^{0.054}$&1.319$_{0.408}^{0.255}$\\
\ MMT07 & 0.3896& 238.0&  19.84 & 20.84&  2.25    &153.87     &  0.00  &0.00 &   0.102$_{0.129}^{0.166}$ &0.872$_{0.007}^{0.008}$&1.777$_{0.123}^{0.132}$\\
\ MMT08 & 0.6335& 239.4&  27.33 & 10.90&  1.88    & 42.30     &  0.00  &0.00 &   0.613$_{0.269}^{0.355}$ &0.584$_{0.096}^{0.100}$&1.260$_{0.411}^{0.332}$\\
\ MMT09 & 0.4788& 235.2&  16.74 &  6.82&  1.49    & 39.48     &  0.00  &0.00 &   0.585$_{0.121}^{0.127}$ &0.762$_{0.086}^{0.093}$&1.338$_{0.285}^{0.261}$\\
\ MMT10 & 0.0683& 286.1&  22.46 & 18.19&  2.02    &121.87     & 59.04  &1.26 &   0.144$_{0.056}^{0.044}$ &0.824$_{0.021}^{0.018}$&1.717$_{0.335}^{0.210}$\\
\ MMT11 & 0.1752& 276.9&  42.42 & 22.01&  2.05    &144.89     & 78.37  &3.13 &   0.386$_{0.034}^{0.039}$ &0.823$_{0.021}^{0.024}$&1.803$_{0.209}^{0.157}$\\
\ MMT12 & 0.6405& 214.5&  23.57 &  6.18&  1.31    & 40.13     &  0.00  &0.00 &   0.840$_{0.360}^{0.601}$ &0.812$_{0.119}^{0.131}$&1.318$_{0.542}^{0.439}$\\
\ MMT13 & 0.4696& 209.1&   6.03 &  6.51&  1.15    & 39.59     &  0.00  &0.00 &   0.072$_{0.160}^{0.231}$ &0.780$_{0.037}^{0.034}$&1.524$_{0.307}^{0.268}$\\
\ MMT14 & 0.4644& 280.6& 100.27 & 56.75&  2.01    &354.12     &  0.00  &0.00 &   0.266$_{0.019}^{0.033}$ &0.794$_{0.005}^{0.005}$&2.215$_{0.150}^{0.112}$\\
\ Keck1 & 0.8390& 124.6&   4.62 &  7.80 & 0.80    & 61.10     &  0.00  &0.00 &  -0.016$_{0.081}^{0.060}$ &0.897$_{0.010}^{0.011}$&1.811$_{0.094}^{0.091}$\\
\ Keck2 & 0.6230&  89.5&   3.49 &  2.33 & 0.67    & 17.60     &  0.00  &0.00 &   0.144$_{0.040}^{0.041}$ &0.843$_{0.021}^{0.019}$&1.440$_{0.148}^{0.134}$\\
\ Keck3 & 0.7906& 121.2&  27.88 & 12.50 & 0.27    & 59.17     &  0.00  &0.00 &   0.658$_{0.042}^{0.038}$ &0.676$_{0.007}^{0.008}$&2.142$_{0.224}^{0.162}$\\
\ Keck4 & 0.8829& 178.5&  13.77 &  7.28 & 0.33    & 35.60     &  0.00  &0.00 &   0.275$_{0.017}^{0.017}$ &0.683$_{0.014}^{0.014}$&2.018$_{0.290}^{0.235}$\\
\ Keck5 & 0.8353& 101.1&   1.85 &  1.10 & 0.58    &  7.99     &  0.00  &0.00 &   0.362$_{0.163}^{0.216}$ &0.879$_{0.035}^{0.035}$&1.102$_{0.367}^{0.252}$\\
\ Keck6 & 0.8237&  90.5&   0.05 &  0.88 & 0.19    &  5.36     &  0.00  &0.00 &  -1.016$_{0.937}^{0.595}$ &0.770$_{0.030}^{0.031}$&1.346$_{0.320}^{0.231}$\\ \hline
\end{tabular}}

Note : the line fluxes are in 10$^{-17}$ \erg with 68 \% confidence uncertainties

\label{tab10}
\end{table*}

\begin{table*}
\centering
\caption{Modelling  the  corrected line ratios to \Hb=1 observed  by  Ly et al. (2014)}
\tiny{
\begin{tabular}{lccccccccccc} \hline  \hline
\  ID          &  OVI  & OV  &  \Ly    &   NV  & CIV  & HeII  &  [OII] & [OIII] & [OIII]  \\
\              &  1034 &1215 & 1215    & 1240  & 1549 & 1640  & 3727+  &  4363  & 5007+       \\ \hline
\ MMT01        &   -   &  -  &    -    &  -    &  -   &  -    & 1.53   &  0.26  & 7.69         \\
\  m1          & 38.7  &7.86 & 25.89   &  3.   & 7.82 & 0.26  & 1.53   &  0.26  & 7.69         \\
\  mpl1        & 5.3   &1.11 & 32.     & 0.43  & 0.68 & 1.45  & 1.5    &  0.05  & 7.8\\
\ MMT02        &   -   &  -  &    -    &  -    &  -   &  -    &4.18    &  0.124 & 4.4     \\
\  m2          & 57.8  & 10.6& 26.2    & 3.75  & 9.7  & 0.14  & 3.9    & 0.23   & 4.2    \\
\ mpl2         & 5.    & 1.  & 30.9    &  0.39 & 0.98 & 1.44  & 4.     & 0.035  & 4.74   \\
\  MMT03       &  -    &  -  &   -     &  -    &   -  &  -    & 5.43   & 0.18   & 2.18   \\
\  m3          & 54.9  & 10.9& 36.7    & 4.2   & 10.9 & 0.117 & 5.2    & 0.2    & 2.5   \\
\  mpl3        & 10.3  & 1.93& 30.8    & 0.65  &1.64  & 1.15  & 5.     & 0.04   & 2.4   \\
\ MMT04        &  -    &  -  &   -     &  -    & -    & -     & 2.26   & 0.07   & 5.94  \\
\   m4         & 46.8  & 9.  &  26.    & 3.65  &9.28  & 0.18  & 2.7    & 0.2    & 5.43  \\
\ mpl4         & 2.7   & 0.6 & 30.7    & 0.32  &0.79  & 1.78  & 3.     & 0.034  & 5.8   \\
\ MMT05        & -     &  -  &  -      &  -    &  -   &  -    & 0.54   & 0.18   & 8.33  \\
\   m5         & 3.16  & 0.67& 24.6    & 0.37  &0.96  & 0.48  & 0.9    & 0.04   & 8.33  \\
\  mpl5        & 2.42  & 0.52& 32.2    & 0.36  &0.91  & 2.11  & 1.2    & 0.07   & 8.4   \\
\ MMT06        &  -    &  -  &  -      &  -    & -    &  -    & 5.42   & 0.167  & 3.49  \\
\  m6          & 69.4  & 16. & 49.7    &  13.4 & 30   &0.425  & 6.     & 0.25   & 3.1   \\
\  mpl6        & 8.3   & 1.52 & 30.5   & 0.51  & 1.31 &1.29   & 5.1    & 0.038  & 3.5   \\
\ MMT07        &    -  &   -  &    -   &  -    &  -   &  -    & 1.26   & 0.12   & 7.44  \\
\  m7          & 36.7  & 7.74 & 26.    & 3.45  & 8.73 & 0.3   & 1.35   & 0.19   & 7.33  \\
\  mpl7        & 5.3   & 1.11 & 32.    & 0.43  & 0.68 & 1.65  & 1.4    & 0.05   & 7.8   \\
\  MMT08       &   -   &  -   &  -     &  -    &  -   &  -    & 4.1    & 0.21   & 3.84  \\
\  m8          & 57.5  & 10.5 & 26.    & 3.73  & 9.61 & 0.14  & 4.     & 0.226  & 3.94  \\
\ mpl8         & 6.65  & 1.3  & 30.9   & 0.5   &1.25  &1.28   & 4.4    & 0.036  & 3.88\\
\ MMT09        &  -    &  -   &  -     &  -    & -    &  -    & 3.84   & 0.265  & 5.78 \\
\  m9          & 59.2  & 11.1 & 26.2   & 4.2   & 10.8 & 0.18  & 3.5    & 0.264  & 5.9   \\
\ mpl9         & 4.1   & 0.83 & 31.2   & 0.37  & 0.92 & 1.66  & 3.71   & 0.04   & 5.75  \\
\ MMT10        &  -    &  -   &  -     &  -    & -    &  -    & 1.39   & 0.128  & 6.67 \\
\  m10         & 26.8  & 6.35 & 25.9   & 0.86  & 8.75 & 0.23  & 1.41   & 0.16   & 6.42 \\
\ mpl10        & 7.    & 1.38 &  32.3  & 0.54  &0.58  & 1.54  & 1.3    & 0.04   & 6.5  \\
\ MMT11        &  -    &  -   &  -     &  -    & -    &  -    & 2.43   & 0.1    & 6.65 \\
\  m11         & 30.5  & 6.63 & 25.6   & 0.63  & 6.75 & 0.123 & 2.     & 0.167  & 6.0  \\
\ mpl11        & 2.66  & 0.52 & 32.8   & 0.21  &0.53  & 1.43  & 2.3    & 0.04   & 6.3  \\
\ MMT12        & -     &  -   &  -     &  -    & -    &  -    & 6.9    & 0.31   & 6.5  \\
\ m12          & 118.7 & 21.8 & 27.8   & 7.7   & 19.8 &0.27   & 6.7    & 0.45   & 6.2  \\
\ mpl12        & 57.6  & 20.4 & 27.1   & 3.    & 8.2  & 1.17  & 6.2    & 0.28   & 6.4  \\
\  MMT13       & -     &  -   &  -     &  -    & -    &  -    & 1.18   & 0.18   & 6.03 \\
\  m13         & 26.   & 6,25 & 25.9   & 0.9   & 9.12 & 0.21  & 1.39   & 0.15   & 6.0  \\
\ mpl13        & 4.74  & 1.04 & 32.3   & 0.47  &1.23  &1.53   & 1.2    & 0.055  & 6.28 \\
\ MMT14        &  -    &  -   &   -    &  -    &  -   &  -    & 1.85   & 0.04   & 6.22 \\
\  m14         & 72.4  & 17.  & 28.3   & 3.12  & 30.3 & 0.38  & 2.4    & 0.2    & 6.1  \\
\ mpl14        & 4.64  & 1.17 & 26.45  & 1.47  & 3.3  & 2.55  & 1.8    & 0.059  & 6.1  \\ 
\ Keck1        & -     &  -   &   -    &  -    &  -   &  -    & 0.96   & 0.12   & 7.89  \\
\ m15          & 28.9  & 6.3  & 25.8   & 2.8   &7.25  & 0.47  & 0.9    & 0.16   & 7.8   \\
\ mpl15        & 2.4   & 0.52 & 32.4   & 0.36  &0.9   & 2.02  & 1.2    & 0.063  & 7.88 \\
\ Keck2        &  -    &   -  &   -    &  -    & -    & -     & 1.39   & 0.25   & 6.97 \\
\ m16          & 0.    & 0.02 & 24.3   & 0.043 & 10.9 & 0.165 & 1.3    & 0.29   & 6.7  \\
\ mpl16        & 1.41  & 0.33 & 34.9   & 0.23  & 0.59 & 1.96  & 1.27   & 0.05   & 7.  \\
\ Keck3        & -     &  -   &   -    &  -    &  -   &   -   & 4.55   & 0.034  & 4.74 \\
\ m17          & 13.6  & 10.4 & 27.4   & 4.12  & 12.9 & 0.07  & 4.     & 0.32   & 4.57 \\
\ mpl17        & 0.081 & 0.3  & 26.4   & 0.42  & 1.81 & 1.55  & 4.4    & 0.049  & 4.75 \\
\ Keck4        & -     &  -   &   -    &   -   & -    & -     & 1.88   & 0.046  & 4.82 \\
\ m18          & 10.7  & 2.2  & 24.9   & 0.68  & 1.78 &0.047  & 1.8    & 0.057  & 4.8 \\
\ mpl18        & 2.45  & 0.54 & 35.    & 0.3   &0.72  &2.57   & 1.6    & 0.035  & 5.0 \\
\ Keck5        &   -   &  -   &   -    &  -    & -    &  -    & 2.3    & 0.6    & 7.56 \\
\ m19          & 0.003 & 0.3  & 27.3   & 0.74  & 20.7 & 0.15  & 2.5    & 0.45   & 8.   \\
\ mpl19        & 0.03  & 0.095& 34.4   & 0.14  & 0.64 &1.95   & 2.1    & 0.05   & 7.6  \\
\ Keck20       & -     &   -  &   -    &   -   & -    &  -    & 0.096  & 0.27   & 5.9 \\
\ m20          &0.     & 0.03 & 23.8   & 0.07  & 14.0 & 3.72  & 0.11   & 0.15   & 6.4  \\
\ mpl20        &420.   & 190. & 38.    & 85.   & 22.5 & 7.9   & 0.08   & 0.43   & 6.  \\ \hline\\

\end{tabular}}

\end{table*}

The R$_{[OIII]}$ ([OIII]5007+/[OIII]4363) line ratios  
 indicate that  gas densities and temperatures in  ranges larger than those
deduced in average from the observations should be accounted for.
This is valid for  the AGN NLR  spectra  (see e.g. (Contini \& Aldrovandi 1986, fig. 4 and references therein, etc.)
and for most of the objects throughout the galactic medium (such as SNR, novae, symbiotic stars) and
the  extra-galactic one.
In Seyfert type 1 galaxies  [OIII] lines  from the BLR  are collisionally deexcited. 
In Seyfert type 2  the [OIII] lines correspond
to gas at \Vs=300-500 \kms and densities of 10$^4$-10$^5$ \cm3.  In intermediate type galaxies
 \RO3 ranges between high ($>$ 100) and low values
($<$ 20).  
  LINERs  (Heckman 1980, Ferland \& Netzer 1983, Ho et al 1993)
cannot be considered as a low $F$ case (Contini \& Aldrovandi 1983, fig. 3).
Problems  can be solved by adding  shock wave hydrodynamics to the calculations of the spectra (Contini 1997 and references therein)
 or by assuming a density gradient  or a  stratification of emitting filaments (Pequignot 1984, Filippenko 1984, etc.)
The importance of the [OIII]4363 line recently emerged for high z galaxies,   because constraining  the models.
To  cross-check this issue we calculate the \Ly/\Hb line ratios for galaxies which show spectra constrained by the
[OIII] 4363 line (see Contini 2014a, Ly et al.2014).
We choose the Ly et al (2014) survey because the  reddening corrected spectra contain  [OIII]5007+4959,
[OIII] 4363, [OII] 3727, \Hb and \Ha lines.

In Table 10  the results of  Ly et al. (2014) observations obtained by the Subaru telescope are reported.
We compare the calculated with observed line ratios in Table 11.
 The  uncertainty for  model calculations is about 10 \%.

In Table 12   models m1-m20  are calculated by  SB dominated (+shock) models,
while models mpl1-mpl20 (described in Table 13) are calculated  by  AGN (+shocks) dominated models. 
Table 11 show that  [OII]/\Hb and [OIII]/\Hb line ratios
are well fitted by both  SB and AGN models, but, regarding the
[OIII]5007+/[OIII]4363  line ratios, 
the  SB dominated models fit the data of nearly all galaxies,
 while MMT5, MMT12, and Keck3 are better reproduced by  AGN dominated
models. For MMT4, Keck4 and Keck6 the  SB and AGN photoionizing models
act  with similar weights.

The [OIII]4363/\Hb is the key line ratio for choosing between  SB or AGN dominated models.
Indeed the shock plays an important role, but the effect of photoionization cannot be neglected.
The ionization parameter ($U$) and the SB effective temperature (\Ts) for  SB and
the power-law flux from the AGN are determined by fitting the [OIII]5007+/\Hb and [OII]3727/\Hb
line ratios. With those input parameters, the [OIII]4363/\Hb line ratios for  SB are  higher
by  factors  very roughly $\sim$ 10 than those calculated for the AGNs, indicating that the shock effect is
relatively strong. So [OIII]5007+/[OIII]4363 are  lower for starbursts.

 The densities in the downstream emitting regions are
higher by a factor of $\sim$ 10, depending on \Vs.
Regarding the [NII]/\Hb line ratios, 
the \Ha/\Hb line ratios  for MMT10 and MMT11  are roughly  between 3 and 4,
the [NII]/\Hb line ratios were directly calculated.
The result of N/H is given only for the  SB dominated models
which    well  reproduce all the other line ratios.
We obtained low N/H relative abundances (0.25 10$^{-4}$)
for the two galaxies MMT10 and MMT11.
The results  for O/H (in 10$^{-4}$ units)  are shown in  Tables 12 and 13.  
 O/H relative abundances calculated by detailed modelling are closer to solar
and higher than those calculated by direct methods.

\begin{figure*}
\centering
\includegraphics[width=16.0cm]{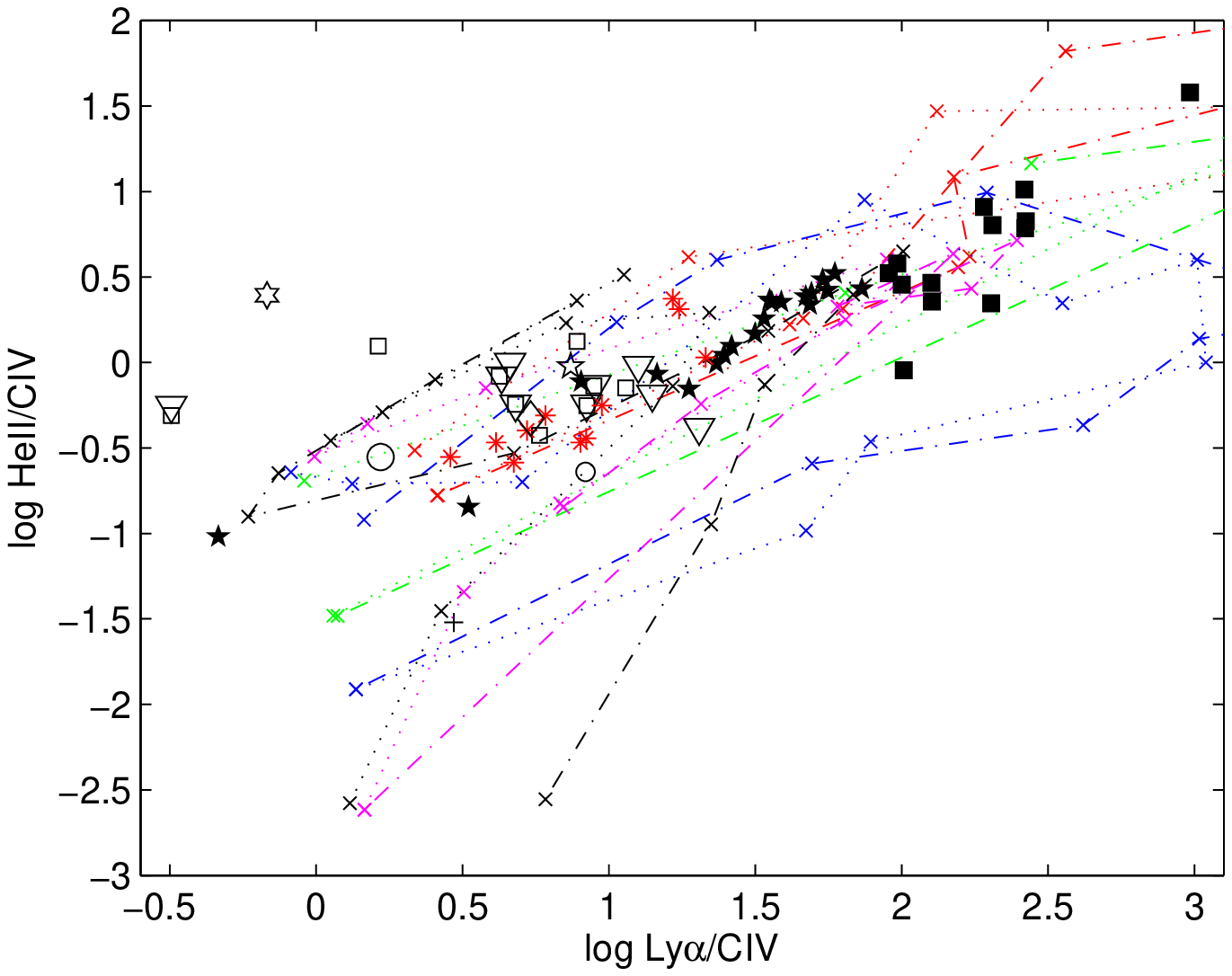}
\caption{The HeII/CIV versus Ly/CIV for NLR  of AGN. Open circle :  Norman et al,
 hexagon : Dawson et al, diamond : Stern et al,  large pentagram : Dey et al;
red asterisks : Collins et al., triangles : Vernet et al, open squares : Humphrey et al.;filled pentagrams
(small for SB and large for AGN models):Ly et al. ; black filled squares  (small for SB and large for AGN models) :
Finkelstein et al.;  small circle: Erb et al ; plus: Fosbury et al.
Line symbols : see Table 1. 
}
\end{figure*}

\section{Discussion and concluding remarks}

 In  previous sections we have been investigating the gas physical conditions and element abundances
in galaxies at relatively high z, selected by their characteristic line spectra, i.e.
 including the \Ly line  and the lines  which constrain the models. 
A detailed modelling process has been used.
All the models are calculated at the emitting nebulae  in the different galaxies.
They lead to \Ly/\Hb higher than observed, except for  some SB dominated models, MDS (Table 4) , MV1 and MV6
(Table 5) and some  models in the Finkelstein et al survey (Tables 7-9).
The  models best reproducing the  data in Tables 2-6 show that 
 most of the spectra  are fitted by  the summed contributions of two types :
 those emitted from the  high 
velocity ($\sim$ 700-1500 \kms) clouds (generally close to an AGN),
which explain the high ionization level lines and UV - X-ray data in the continuum SED,
 and  those emitted from the relatively low \Vs ($\sim$ 100-300 \kms), \n0 (100-300 \cm3) radiation dominated 
 clouds (by a AGN or a SB)
which  confirm  the narrow optical line profiles, as well as  the optical spectral and continuum data. 
 In some cases one or more    model components show that  the photoionizing flux from the active source is  
prevented  from reaching the gas,
or that the  source is absent. 
Therefore, the clouds within the galaxies  are ionized and heated 
only by the shock. In the HDF  galaxy  (Dawson et al)
the high \Vs (1300 \kms) clouds are found in the surroundings of a SB corresponding to a high \Ts.
In the BX418 galaxy (Erb et al), the high velocity gas  (\Vs=600-800 \kms) shows
 a preshock density higher by a factor  $>$ 10 than in  the other objects.
The   shock velocities were most probably reduced by collision with  high density gas.
These clouds are  located in the surroundings of an AGN.
This is valid also for the shock dominated clouds (model MF0, Table 6) in the Lynx arc (Fosbury et al). 
A few galaxies from the radio survey of Humphrey et al (Table 3) do not show the high \Vs component
which  is neither present in  the galaxies selected from the  
Vernet et al  survey (Table 5). Here, the clouds are characterised by the very low \Vs and \n0, similar 
or lower than those calculated
to  reproduce  the Finkelstein et al spectra (Tables 8 and 9).

We report in Fig. 8   HeII 1640/CIV 1549 versus \Ly/CIV 1549
calculated by the models presented in GRID1, considering
gas in the conditions suitable to the NLR of AGN. 
The line ratios  were calculated by GRID1    adopting two different 
geometrical thickness of the emitting clouds, $D$=10$^{17}$ cm and 10$^{19}$ cm. 
In GRID1, for each set  of \n0 and \Vs   a series of models with different $F$, (log $F$=8,9,10,11,12) were adopted.
$F$ =0 represents the shock dominated  case and corresponds to the lowest HeII/CIV and \Ly/CIV.
Solar (Allen 1976)  element abundances were  used  to calculate the  spectra.
 The  HeII/CIV and   \Ly/CIV trends (for each \Vs) follow the  increasing  photoionization flux,  because
both HeII and \Ly are recombination lines which depend on  photoionization,   while 
 the   fractional abundance distribution of  carbon ions in different ionization levels   strongly depends on  gas
 temperature and  density  which  are affected by the shock.

The data observed for nearly all the AGN dominated objects   are nested between models showing 
\Vs $\sim$  100 and 200 \kms, the fluxes ranging throughout
 a large scale, and both geometrically thin and thick  clouds.
Also shown in Fig. 8 are the  spectral data  observed from the local galaxy Mrk3 (Collins et al 2005) because 
they    include the   \Ly  and UV lines.
The data (reddening corrected) refer to different regions within the galaxy. 
The spectra  represent  average  conditions
  covering small regions within the galaxy. So  they tend to show   intrinsic  conditions of the gas rather 
than the averaged ones    obtained by  modelling the  observed spectra   covering the entire galaxy, 
as necessarily occurs for high z galaxies.
The Mrk3 data are less scattered than those  observed from high z galaxies and follow  the model trends.

In Fig. 8 we added    the data based on the Finkelstein et al and  Ly et al. observations.
They do not represent   observed HeII/CIV versus observed \Ly/CIV, but the calculated line ratios which
  are strongly  constrained
by  the   observed optical lines.  Finkelstein et al and  Ly et al. objects
follow the trends of AGN (+shock) dominated  models with roughly constant velocities. 
The observed line ratios referring to  the other galaxies
 are scattered  for different $F$.
The  Finkelstein et al and  Ly et al  samples are  situated close to  higher  shock velocity models.

Finally, in Figs. 9  the physical parameters (\Vs, \n0, \Ts, $U$, $F$, $D$), the \Ha flux
 at the nebula and  O/H and N/H relative abundances
calculated for the present  sample of  galaxies are
presented as function  of the  SFR.
  SFR were calculated by the \Ha flux measured at Earth by the observers.
We chosed \Ha instead  of \Ly  because less affected by scattering and absorption
problems.
For galaxies in which the \Ha data were not  available, we used  \Ly  and we calculated
 \Ha line fluxes by noticing that the \Ly/\Hb line ratios can be approximated to $\sim$25 for many models (Fig. 1)
and that \Ha/\Hb $\sim$ 3.

The sample of galaxies which   emit    \Ly  and UV lines   do not  show
peculiar physical conditions of the emitting gas, nor abnormal element abundances.  However, for some galaxies of 
the Humphrey et al, Dawson et al, Stern et al,
Dey et al, Norman et al surveys,  emitting gas with relatively high \Vs  ($>$ 1000 \kms)   is revealed by 
the complex FWHM of the line profiles. 
Such velocities are  upper limits to  those characteristic of the AGN NLR. Generally, higher \n0 accompany
higher \Vs leading to collisionally deexcited forbidden lines. This is not  the case for the \Ly line 
emitting galaxies modelled throughout this paper. 
In the Humphrey et al galaxy 0211-122 spectrum reported in Table 2, for instance, [OIII] 5007 lines 
with FWHM = 1300$\pm$60 \kms and [OII] 3727 with FWHM=730$\pm$160 \kms were observed. 
 Most of the high velocity models are shock dominated. In these cases, the electron temperature and density
steeply drop to T$\geq$10$^3$ K (Fig. 6) at a certain distance from the shock-front (which depends on \Vs)
 because the primary flux is absent and the secondary one
is relatively weak.

The high \Ts ($>$  10$^5$ K) which indicates that the SB stars  are close to outburst (Contini 2014b)
were found  in a few  objects  (4C+23.56 at z=2.479 from the Humphrey et al survey,  SST24 at z=2.858
from the Dey et al survey and   the Lynx arc at z=3.357 from  Fosbury et al observations )
 within more than 50 galaxies in the sample.
These  outstanding temperature, which  are  characteristic of galaxies showing  some activity (Contini 2014b),   
are  isolated events at the state of the art.
Considering the evolutionary paths in the H-R diagrams for stars of different mass during the pre-main-sequence phase,
these temperatures can be reached by $\geq$9\msol stars in a lifetime of $\geq$  10$^5$ years (Prialnik 2000, fig. 8.1), 
which is very short compared to the redshift z  age ($>$ 2 Gyr).

 Fig.  9  shows that the SFR between $\sim$ 1 and $\sim$ 300 \Moy is accompanied by relatively high shock velocities ($>$ 1000 \kms), 
but gaseous cloud densities
are relatively low, similar to those in the AGN NLR and in the SB surrounding (\n0$\leq$350 \cm3)
  for SFR $\geq$ 10 \Moy.
O/H  reaches a lower limit  minimum for some galaxies at  SFR$<$ 10 \Moy, then it grows with SFR to solar values
 at SFR $\sim$ 300 \Moy.
The N/H relative abundances show a large scattering. N/H are particularly low  for a few 
 galaxies in the Finkelstein et al survey at z$\sim$ 0.25.
Humphrey et al, however,  claim that
nuclear activity and star formation which lead to solar metallicities  at early times,
 are seen through lack of variability and high metallicity between z=2.5 and 0.
The geometrical thickness of the emitting clouds presents
 minimum  $D$ $\leq$ 10$^{16}$ cm. 
 $D$ and consequently also  \Ha  fluxes calculated at the nebula
show  a large scattering  covering roughly 4 orders of magnitude for SFR between 1 and 650 \Moy, indicating fragmentation
by turbulence in a shock dominated regime.

\begin{table}
\centering
\caption{The physical parameters and the relative abundances for Ly et al  SB dominated models (m1-m20)}
\tiny{
\begin{tabular}{cccccclcccccc} \hline  \hline
\ model  &    \Vs $^1$  &   \n0 &   \Ts   &    U   &    D        &  O/H   &                 \Hb calc  \\
\       &     1         &   2   &  3      &    -    &   4        &   5     &    6                        \\ \hline
\ m1    &   220   & 120 &    5.           &  0.05  &  8.         &        6.5    &                 0.00176   \\
\ m2    &   260   & 120 &    3.8          &  0.03  &  8.6        &        7.3    &                 0.0016   \\
\ m3    &   240   & 130 &    1.           &  0.001 &  8.         &        6.6   &                 0.0015   \\
\ m4    &   260   & 110 &    4.7          &  0.03  &  10.        &        6.3    &                 0.0016   \\
\ m5    &   250   & 110 &    7.           &  0.06  &  40.        &        4.5    &                 0.018    \\
\ m6    &   260   & 130 &    8.           &  0.0005&  28.        &        3.     &                 0.0008   \\
\ m7    &   220   & 120 &    5.1          &  0.05  &  8.         &        5.6    &                 0.0017   \\
\ m8    &   260   & 120 &    3.7          &  0.03  &  8.6        &        7.3   &                 0.0016   \\
\ m9    &   260   & 110 &    4.4          &  0.03  &  9.5        &        6.8   &                 0.0013   \\
\ m10   &   200   & 120 &    5.           &  0.04  &  8.         &        4.7   &                 0.0018   \\
\ m11   &   200   & 120 &    4.5          &  0.04  &  8.         &        6.6   &                 0.003     \\
\ m12   &   260   & 120 &    3.6          &  0.02  &  7.7        &        7.3   &                 0.00076 \\
\ m13   &   200   & 120 &    4.9          &  0.04  &  8.         &        4.4    &                 0.00176 \\
\ m14   &   200   & 120 &    4.6          &  0.03  &  7.         &        5.5    &                 0.0014  \\
\ m15   &   200   & 120 &    5.4          &  0.06  &  8.         &        5.5    &                 0.0019  \\
\ m16   &   90    & 120 &    4.3          &  0.06  &  5.         &        6.6   &                 0.0004  \\
\ m17  &   140   & 120 &    3.4           &  0.03  &  2.8        &        7.35  &                 0.00064 \\
\ m18   &   200   & 140 &    4.5          &  0.04  &  13.        &        8.    &                 0.0075  \\
\ m19   &   100   & 120 &    4.5          &  0.03  &  3.         &        5.5   &                 0.0003  \\
\ m20   &   90    & 100 &    5.7          &  0.2   &  8.         &        5.5   &                 0.00035 \\ \hline
\end{tabular}}

1: in \kms ; 2: in \cm3; 3: in 10$^4$K; 4: in 10$^{15}$ cm; 5: in 10$^{-4}$; 6: in \erg
\label{tab12}

\centering
\caption{The physical parameters and the relative abundances for Ly et al  AGN dominated models (mpl1-mpl20)}
\tiny{
\begin{tabular}{ccccclccccccccccc} \hline  \hline
\ model  &   \Vs &  \n0  &  F       &  D          &  O/H   &   \Hb calc  \\
\        &   1   &   2   &  3       &   4         &   5        &   6         \\ \hline
\ mpl1   &  220  &  1000 &   8      &   1         &  6.6       &    0.124    \\
\ mpl2   &  220  &  370  &   2      &   10        &  6.6       &    0.048    \\
\ mpl3   &  240  &  330  &   1      &   20        &  7.6       &    0.025    \\
\ mpl4   &  220  &  320  &   2.8    &   30        &  4.6       &    0.06     \\
\ mpl5   &  280  &  620  &   10     &   3         &  3.6       &    0.158    \\
\ mpl6   &  250  &  300  &   1.2    &   15        &  7.6       &    0.03     \\
\ mpl7   &  220  &  1000 &   8      &   1         &  6.6       &    0.124    \\
\ mpl8   &  240  &  330  &   1.5    &   10        &  6.6       &    0.036    \\
\ mpl9   &  240  &  330 &    2.4    &   12        &  5.6       &    0.053    \\
\ mpl10  &  250  &  1000&    7.6    &   1         &  6.6       &    0.11     \\
\ mpl11  &  250  &  550 &    6      &   7         &  6.6       &    0.15    \\
\ mpl12  &  240  &  160 &    0.25   &   5.6       &  8.9       &    0.0021   \\
\ mpl13  &  220  &  1000&    7.4    &   1         &  5.6       &    0.125    \\
\ mpl14  &  250  &  160 &    1      &   12        &  2         &    0.01      \\
\ mpl15  &  220  &  780 &    1.3    &   4         &  3.6       &    0.24     \\
\ mpl16  &  220  &  720 &    12     &   4         &  3.6       &    0.24     \\
\ mpl17  &  120  &  260 &    0.6    &   9         &  5.        &    0.114    \\
\ mpl18  &  230  &  720 &    7.3    &   4         &  4.5       &    0.17      \\
\ mpl19  &  120  &  920 &    7      &   4         &  4.6       &    0.13     \\
\ mpl20  &  90   &  100 &    6      &   2         &  4.6       &    0.00005  \\ \hline
\end{tabular}}

1: in \kms ; 2: in \cm3; 3 in 10 $^{10}$  photons cm$^{-2}$ s$^{-1}$ eV$^{-1}$ at the Lyman limit;
4: in 10$^{15}$ cm; 5: in 10$^{-4}$; 6: in \erg
\label{tab13}
\end{table}


\begin{figure*}
\centering
\includegraphics[width=7.4cm]{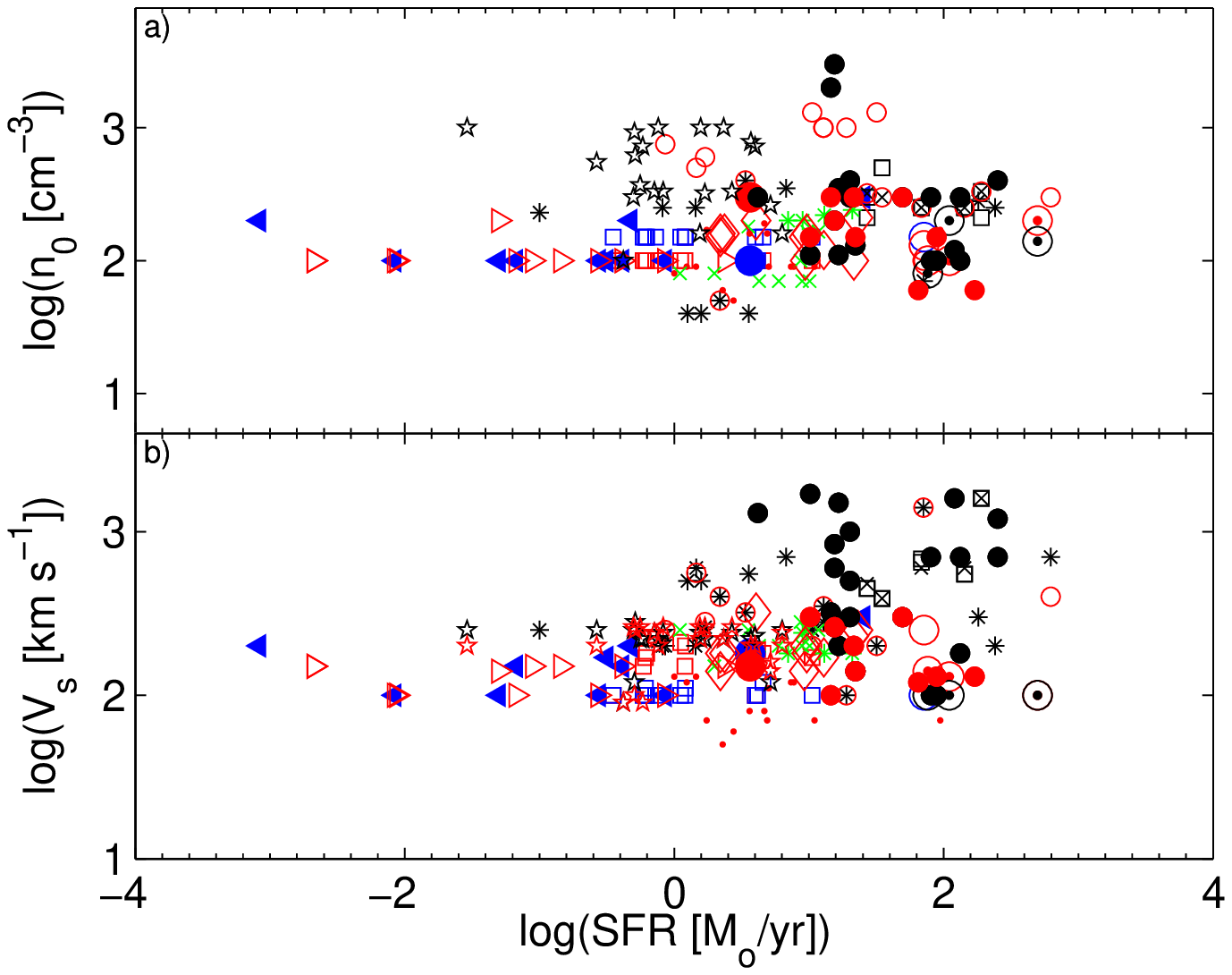}
\includegraphics[width=7.4cm]{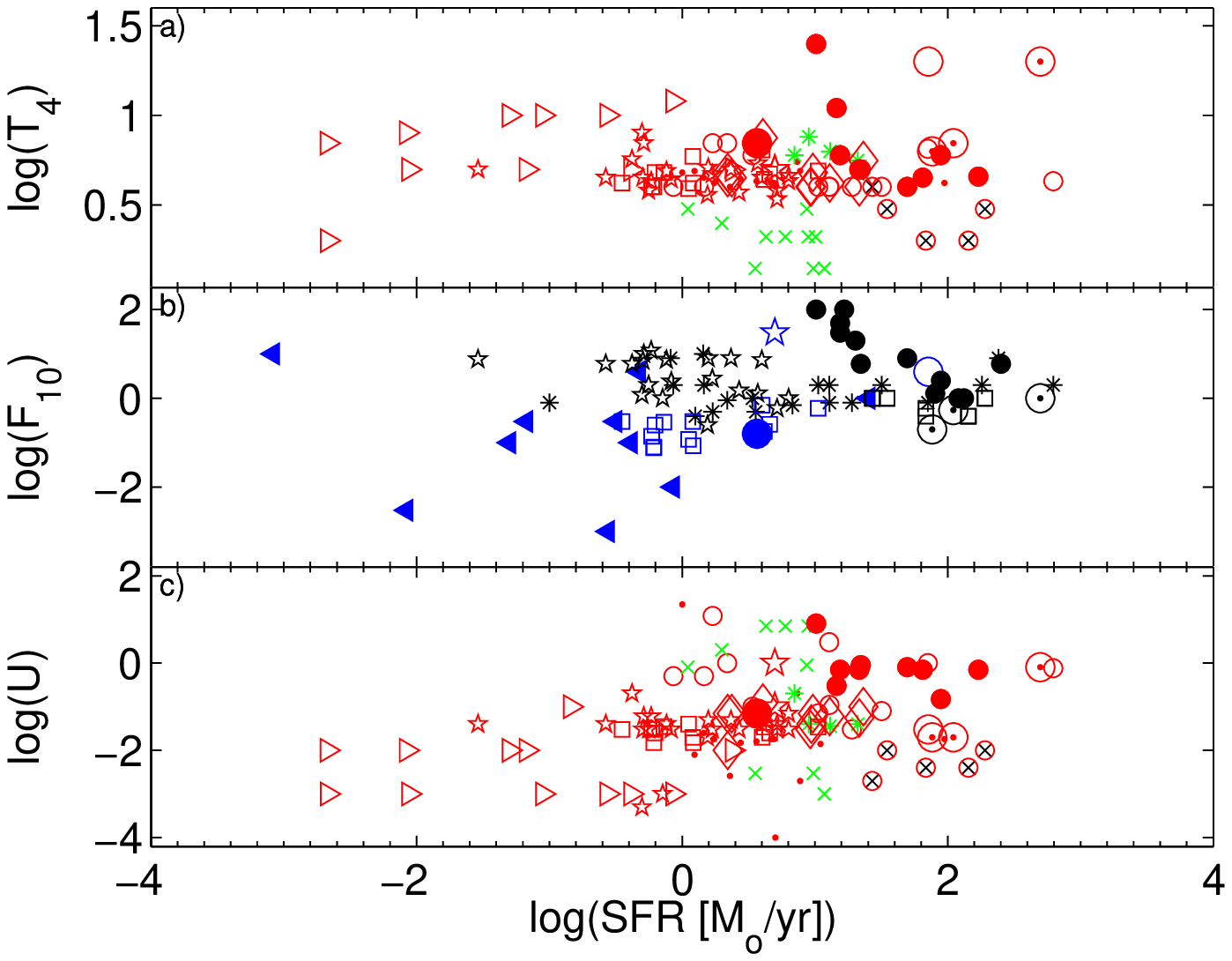}
\includegraphics[width=7.4cm]{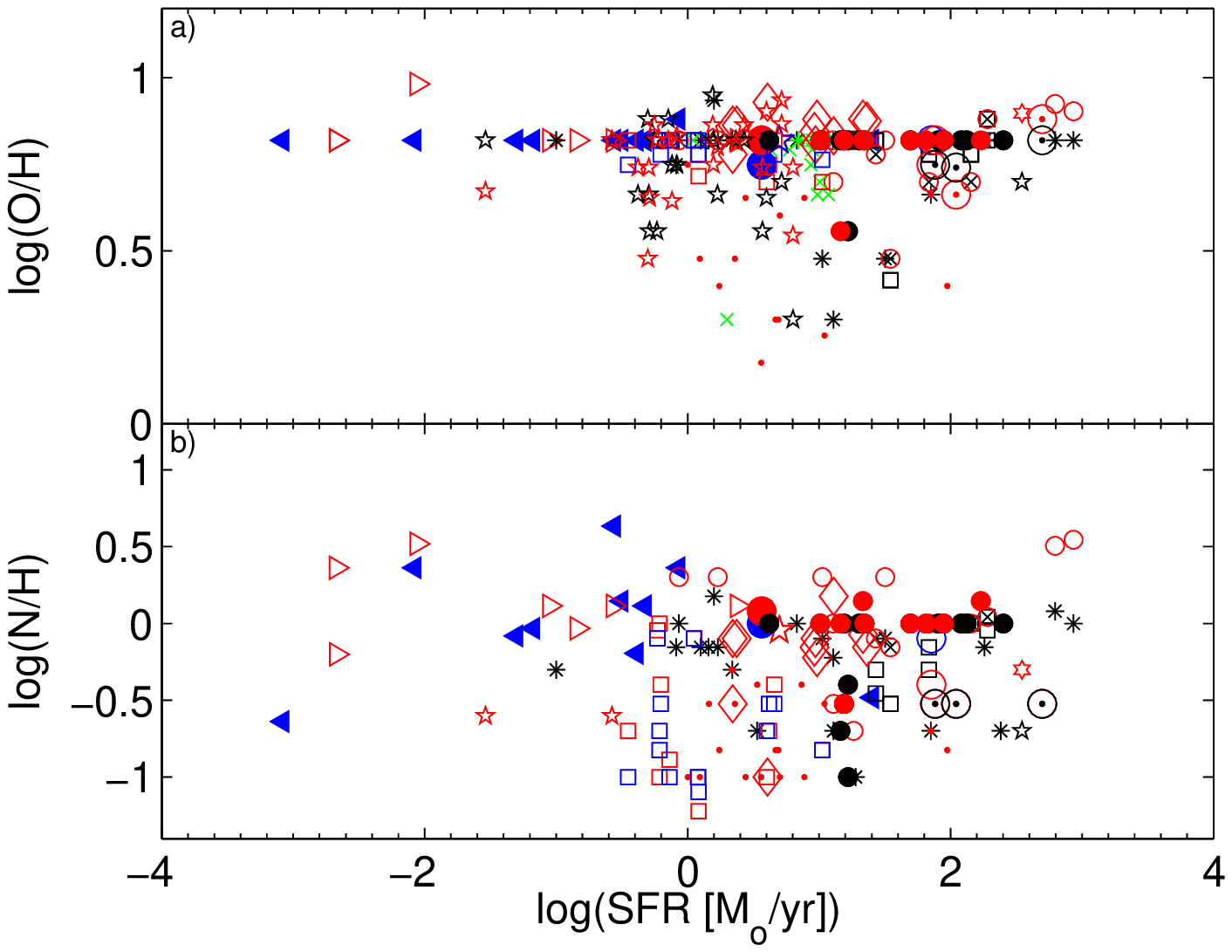}
\includegraphics[width=7.4cm]{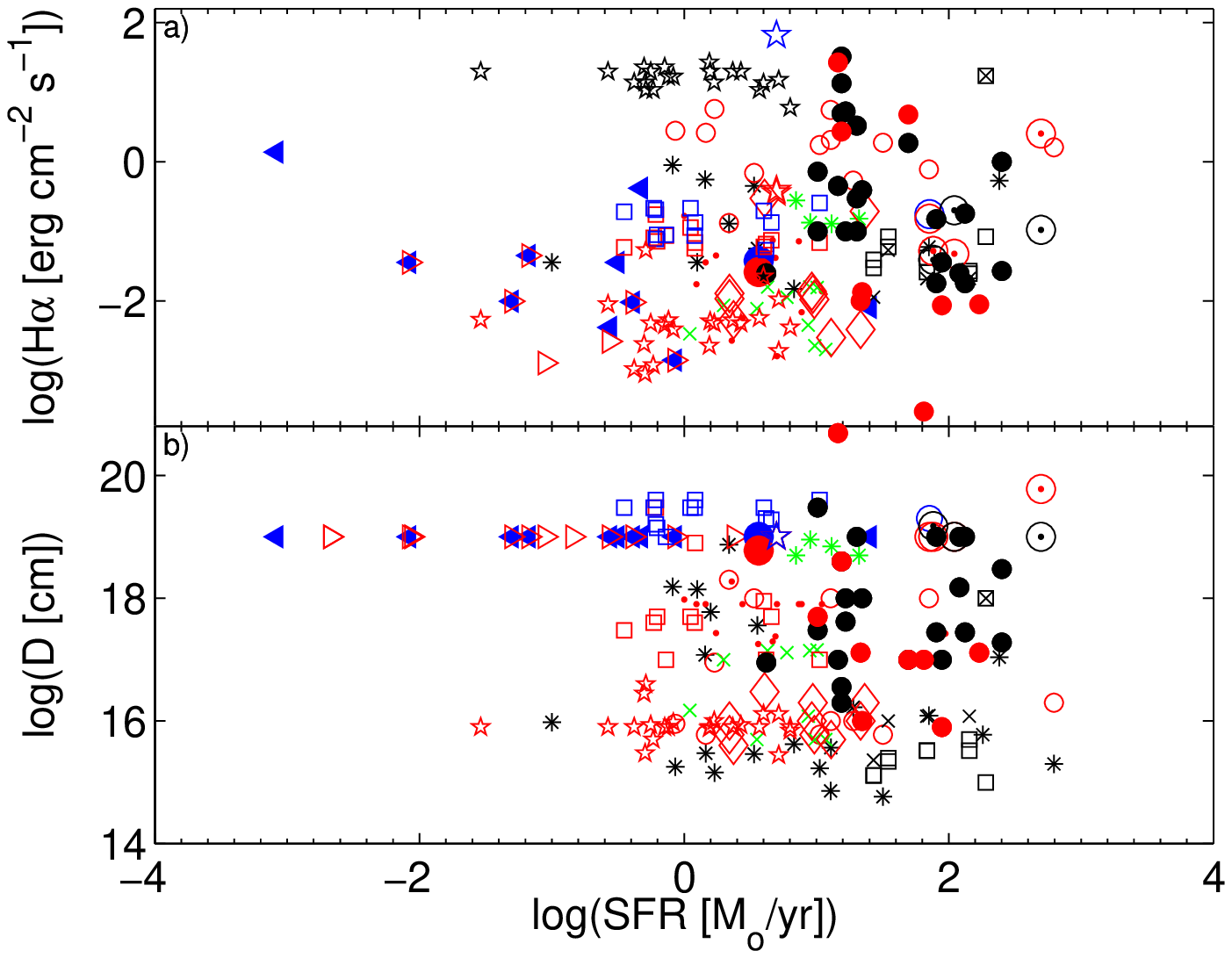}
\caption{The  physical parameters calculated for the galaxy surveys as a function of the SFR.
Symbols as in Contini 2014a, fig. 5. For the present sample :
blue open squares : Finkelstein et al for AGN; red open squares : Finkelstein et al. for SB.
The results for Ly et al (2014) data are represented by pentagrams, red for starbursts and black for AGN;
black filled circles: Dey et al., Dawson et al., Norman et al., Stern et al., Humphrey et al., Vernet et al.,
Erb et al, and Fosbury et al.
AGN models; red filled circles : the same for SB dominated models.}
\end{figure*}

\section*{Acknowledgements}
I am very grateful to the referee for reviewing the paper and for many comments which improved 
its presentation. I  thank Dr. Chun Ly for providing the data in suitable format.

\section*{References}

\def\ref{\par\noindent\hangindent 18pt}

\ref Aldrovandi, S.M.V. \& Contini 1994, A\&A, 140, 368
\ref Allen, C.W. 1976 Astrophysical Quantities, London: Athlone (3rd edition)
\ref Atek, H.; Kunth, D.; Schaerer, D.; Mas-Hesse, J.; Hayes, M.;  Östlin, G.; Kneib, J-P, 2014, A\&A, 561,
89	
\ref Beck, R. 2012,JPhCS, 372, 02051
\ref Bose, S. et al. 2015, arXiv:1504.06207
\ref Charlot,S., Fall, S.M. 1993, ApJ, 415, 580
\ref Cimatti, A., di Serego Alighieri, S., Vernet, J., Cohen, M., Fosbury, R.A.E. 1988, ApJ, 499, L21
\ref Collins, N. R.; Kraemer, S. B.; Crenshaw, D. M.; Ruiz, J.; Deo, R.; Bruhweiler, F. C. 2005, ApJ, 619, 116	
\ref Contini, M. 2014a, A\&A, 564, 19
\ref Contini, M. 2014b, A\&A, 572, 65
\ref Contini, M. 2013a, MNRAS, 429, 242
\ref Contini, M. 2013b, arXiv:1310.5447
\ref Contini, M. 2012, MNRAS, 425,1205
\ref Contini, M. 2009, MNRAS, 399, 1175
\ref Contini, M. 1997, A\&A, 323, 71
\ref Contini, M., Aldrovandi, S.M. 1983, A\&A, 127, 15
\ref Contini, M., Aldrovandi, S.M. 1986, A\&A, 168, 41
\ref Contini, M. \& Viegas, S.M.A. 2001a, ApJS, 132, 211
\ref Contini, M. \& Viegas, S.M.A. 2001b, ApJS, 137, 75
\ref Contini, M., Angeloni, R., Rafanelli, P. 2009, A\&A, 496, 759
\ref Contini, M., Cracco, V., Ciroi, S., \& La Mura, G. 2012, A\&A, 545, A72
\ref Contini, M., Viegas, S.M., Prieto, M.A. 2004, MNRAS, 348, 1065
\ref Cox, D.P. 1972, ApJ, 178, 143
\ref Crenshow, D.M. et al 2002, ApJ, 566, 187
\ref Dawson, S.; McCrady, N.; Stern, D.; Eckart, M. E.; Spinrad, H.; Liu, M. C.; Graham, J. R.  
	 2003, AJ, 125, 1236	
\ref Dey, A. et al. 2005, ApJ, 629, 654
\ref Dwek, E. 1981, ApJ, 247, 614
\ref Erb, D.K., Pettini, M., Shapley, A.E.,  Steidel, C.C., Law, D.R., Naveen, A.R. 2010, ApJ, 719, 1168
\ref Ferland, G., Netzer, H. 1983, ApJ, 264, 105 
\ref Field, G.B., Steigman, G. 1971, ApJ, 166, 59
\ref Filippenko, A.V. 1985, ApJ, 289, 475
\ref Finkelstein, S. L.; Cohen, S. H.; Moustakas, J.; Malhotra, S., Rhoads, J.E.; Papovich, C. 2011, ApJ, 733, 117	
\ref Fosbury, R.A.E. et al 2003, ApJ, 596, 797
\ref Hamann, F. \& Ferland, G. 1993, ApJ, 418, 11
\ref Heckman, T.M. 1980 A\&A, 87, 152
\ref Heng, K., Sunyaev, R.A. 2008, A\&A, 481, 117
\ref Ho,L.C., Filippenko, A.V., Sargent, W.I.W. 1993, Ap], 417, 63
\ref Humphrey, A.; Villar-Martin, M.; Vernet, J.; Fosbury, R.; di Serego Alighieri, S.; Binette, L.
2008, MNRAS, 383, 11	
\ref Knopp, G.P. \& Chambers, K.C. 1997, ApJS, 109, 367
\ref  Ly, C.; Malkan, M. A.; Nagao, T.; Kashikawa, N.; Shimasaku, K.; Hayashi, M.  2014, ApJS, 780, 122
\ref Norman, C. et al 2002, ApJ, 571, 218
\ref Osterbrock, D.E.  Astrophysics of Gaseous Nebulae and Active Galactic Nuclei. Mill Valley, CA:
University Science Books; 1989.
\ref Oteo, I. et al. 2012, A\&A, 541, 650
\ref Pequignot, D. 1984, A\&A, 131, 159
\ref Prialnik, D. 2000 in An Introduction to the Theory of Stellar Structure and Evolution, Cambridge University Press
\ref Richards, E.A. 2000, ApJ, 533, 611
\ref Rodr\'{i}guez-Ardila,A., Contini, M., Viegas, S.M. 2005, MNRAS, 357, 220
\ref Ryan Jr.,R.E., Cohen, S.H., Windhorst, R.A., Silk, J. 2008, ApJ, 678, 751
\ref Stanford, S.A.,Elston, R., Eisenhardt,P.,R.,M., Spinrad,H., Stern, D., Dey, A. 1997, AJ, 114, 2232
\ref Steigman, G., Werner, M.W., Geldon, F.M. 1971, ApJ, 168, 373
\ref Stern, D. et al. 2002, ApJ, 568, 71
\ref Turner, T.J. et al 2001, ApJ, 561, 131
\ref Vernet, J.; Fosbury, R. A. E.; Villar-Martin, M.; Cohen, M. H.; Cimatti, A.; di Serego Alighieri, S.; Goodrich, R. W. 
2001, A\&A, 366, 7	
\ref  White, H.E., 1934,  Introduction to Atomic Spectra, McGROW- HILL book company, New York and London.
\ref Williams, R.E. 1967, ApJ, 147, 552
\ref Yajima, H., Li, Y., Zhu, Q. 2012 arXiv:1210.6440

\end{document}